\documentstyle[12pt]{article}
\def\ie{{\em i.e.}}
\def\eg{{\em e.g.}}

\def\beq{\begin{equation}}
\def\eeq{\end{equation}}

\catcode`\@=11 
\def\coeff#1#2{{\textstyle{#1\over #2}}}

\def\VEV#1{\left\langle #1\right\rangle}
\def\vev#1{\left\langle #1\right\rangle}
\def\lsim{\mathrel{\mathpalette\@versim<}}
\def\gsim{\mathrel{\mathpalette\@versim>}}
\def\@versim#1#2{\vcenter{\offinterlineskip
    \ialign{$\m@th#1\hfil##\hfil$\crcr#2\crcr\sim\crcr } }}
\def\etal{{\em et. al.}}
\def\JL{J. L. Lopez}
\def\DVN{D. V. Nanopoulos}
\def\AZ{A. Zichichi}

\def\r#1{$\bf#1$}
\def\rb#1{$\bf\overline{#1}$}

\def\t1{{\tilde 1}}

\def\F{\widetilde F}
\def\Fb{\widetilde{\bar F}}

\def\GeV{\,{\rm GeV}}
\def\TeV{\,{\rm TeV}}

\def\to{\rightarrow}

\def\b{(T+\bar T)}

\def\half{\coeff{1}{2}}

\def\NPB#1#2#3{Nucl. Phys. B {\bf#1} (19#2) #3}
\def\PLB#1#2#3{Phys. Lett. B {\bf#1} (19#2) #3}

\def\PRD#1#2#3{Phys. Rev. D {\bf#1} (19#2) #3}

\def\TAMU#1{Texas A \& M University preprint CTP-TAMU-#1}

\begin{document}
\begin{flushright}
\baselineskip=12pt
{CERN-TH.7519/94}\\
{CTP-TAMU-60/94}\\
{ACT-18/94}\\
{hep-ph/9412332}\\
Expanded version
\end{flushright}

\begin{center}
\vglue 1cm
{\Huge\bf String no-scale supergravity\\}
\vglue 1cm
{JORGE L. LOPEZ$^{1,2}$ and D.V. NANOPOULOS$^{1,2,3}$\\}
\vglue 0.5cm
{\em $^1$Center for Theoretical Physics, Department of Physics, Texas A\&M
University\\}
{\em College Station, TX 77843--4242, USA\\}
{\em $^2$Astroparticle Physics Group, Houston Advanced Research Center
(HARC)\\}
{\em The Mitchell Campus, The Woodlands, TX 77381, USA\\}
{\em $^3$CERN Theory Division, 1211 Geneva 23, Switzerland\\}
\end{center}

\vglue 1cm
\begin{abstract}
We explore the postulates of string no-scale supergravity in the context
of free-fermionic string models. The requirements of vanishing vacuum energy,
flat directions of the scalar potential, and stable no-scale mechanism impose
strong restrictions on possible string no-scale models, which must possess only
two or three moduli, and a constrained massless spectrum. The
soft-supersymmetry-breaking parameters involving all twisted and untwisted
fields are given explicitly. This class of models contain no free parameters,
\ie, in principle all supersymmetric particle masses and interactions are
completely determined. A computerized search for free-fermionic models with the
desired properties yields a candidate $SU(5)\times U(1)$ model containing extra
(\r{10},\rb{10}) matter representations that allow gauge coupling unification
at the string scale. Our candidate model possesses a bening non-universal
assignment of supersymmetry breaking scalar masses, which may have interesting
low-energy experimental consequences.
\end{abstract}
\vspace{0.5cm}
\begin{flushleft}
\baselineskip=12pt
{CERN-TH.7519/94}\\
{CTP-TAMU-60/94}\\
{ACT-18/94}\\
February 1995
\end{flushleft}

\vfill\eject
\setcounter{page}{1}
\pagestyle{plain}
\baselineskip=14pt

\section{Introduction}
\label{Introduction}
Experiments at LEP and the Tevatron have given the strongest to-date support to
the Standard Model of the strong and electroweak interactions. Yet despite all
this experimental evidence, physicists believe that this model is incomplete.
One possible completion of the Standard Model is embedded in the physics of
supersymmetry. In fact, the same experimental evidence confirming the Standard
Model, can also be used to support its supersymmetric extension. This can be
seen through the unification of the gauge couplings at very high energies
in supersymmetric models, but perhaps more pervasively from the fact that
supersymmetric models have built-in mechanisms that make them look almost
identical to the Standard Model at presently available facilities. This extreme
similarity with the Standard Model occurs at the tree-level for energies below
the threshold for production of supersymmetric particles, and also at
one-loop if the supersymmetric mass scales exceed the electroweak scale. This
similarity is not totally devoid of predictivity, since in supersymmetric
models the lightest Higgs boson is expected to be light ($m_h\sim M_Z$). In
view of these facts, attention has turned strongly towards supersymmetric
models, in particular those that can be understood as low energy limits of more
fundamental theories, such as grand unification, supergravity, and
superstrings. These models are highly predictive, with all supersymmetric
physics typically dependent on four or less parameters.

Four-parameter supersymmetric models are obtained as low-energy effective
supergravity models with assumed universal soft-supersymmetry-breaking terms:
the scalar mass ($m_0$), the gaugino mass ($m_{1/2}$), the trilinear scalar
coupling ($A$), and (at low energies) the ratio of Higgs vacuum expectation
values ($\tan\beta$). Most phenomenological analyses are content with exploring
such a four-dimensional parameter space. However, theoretically speaking it
is not clear that all possible combinations of these parameters are consistent,
since in a specific supergravity model, \ie, one specified by the K\"ahler
function $G$ and the gauge kinetic function $f$, the parameters $m_0,m_{1/2},A$
can be explicitly calculated in terms of the gravitino mass ($m_{3/2}$).
Guidance in this matter has come from string-inspired choices for $G$ and $f$,
which lead to simple values for the ratios $m_0/m_{1/2}$ and $A/m_{1/2}$, and
thus to two-parameter supersymmetric models.\footnote{Note however that such
parametrization may be inadequate since non-universal scalar masses are not
uncommon in string-derived supergravities.}

Despite this great reduction in the model parameters, several questions remain
unanswered: (i) can one construct an explicit string-derived model where the
various ratios of soft supersymmetry breaking terms are calculated, and at the
same time the usual low-energy phenomenology is explained? (ii) does this model
possess a sufficiently suppressed cosmological constant? (iii) how is the scale
of supersymmetry breaking ($m_{3/2}$) determined in such model? This model
would be a ``no-parameter" model.

No-scale supergravity \cite{LN} provides satisfactory answers to the latter two
questions, \ie, vanishing cosmological constant (at the tree level) and
dynamical determination of $m_{3/2}$ via the no-scale mechanism. {\em String
no-scale supergravity} is postulated to be the subset of string models which
can provide satisfactory  answers to all three questions. This subset is rather
restricted, since in practice it is seen that most known string models {\em do
not} obey the postulates of string no-scale supergravity.

Our purpose in the present paper is to explore the postulates of string
no-scale supergravity in the context of fermionic string models. Our
investigations lead to a set of constraints on the spectrum and interactions
in realistic string models, as well as to novel predictions for the
soft-supersymmetry-breaking parameters. We are also able to provide an
existence proof that realistic string no-scale supergravity models do exist. A
search for models of this type with the gauge group $SU(5)\times U(1)$ turns up
a rather interesting phenomenon: among the class of models which we have
explored, a necessary condition to satisfy the postulates of no-scale
supergravity appears to be the existence in the spectrum of extra
(\r{10},\rb{10}) matter representations that allow gauge coupling unification
at the string scale.

This paper is organized as follows. In Sec.~\ref{No-scale} we summarize the
postulates of no-scale supergravity and the no-scale mechanism. In
Sec.~\ref{Fermionic} we discuss the K\"ahler potential and the superpotential
in free-fermionic string models, and compute the vacuum energy, and the
quantity ${\rm Str}\,{\cal M}^2$. We also discuss the conditions under which
these quantities would vanish, as required in no-scale models. In
Sec.~\ref{Soft} we compute the soft-supersymmetry-breaking parameters,
including the Goldstino composition, the gaugino and scalar masses and the $A$
terms. We also discuss the origin of the $\mu$ term and the associated
parameter $B$. In Sec.~\ref{Field} we discuss the normalization of the fields
and how these affect the observable Yukawa couplings. In Sec.~\ref{Possible} we
perform a search for realistic string no-scale free-fermionic models, and
present evidence for the conjecture mentioned in the previous paragraph. We
also study the supersymmetry-breaking parameters in the candidate model found.
Finally, in Sec.~\ref{Conclusions} we summarize our conclusions, in
Appendix~\ref{Redefinitions} we collect some details about the transformation
between the string and supergravity bases, and in Appendix~\ref{AppB} we give
details of the calculation of the twisted sector K\"ahler potential in a
specific model.

\section{No-scale supergravity}
\label{No-scale}
A supergravity theory is specified by two functions, the K\"ahler function
\begin{equation}
G=K+\ln|W|^2\ ,
\label{Gdef}
\end{equation}
where $K$ is the K\"ahler potential and $W$ the superpotential, and the
gauge kinetic function $f$. All masses and interactions are explicitly
calculable from these inputs. In particular, the (tree-level) scalar potential
is given by
\begin{equation}
V=e^G(G^IG_I-3)+V_D\ ,
\label{Vdef}
\end{equation}
where the sum is over all scalar fields in the spectrum, $G_I=\partial_I G$,
$G^I=G^{I\bar J}G_{\bar J}$, and $G^{I\bar J}$ is the inverse K\"ahler metric
(\ie, the transpose of the inverse of $G_{I\bar J}=K_{I\bar J}$). The second
term in Eq.~(\ref{Vdef}) is the contribution from the $D$-terms; we will assume
in what follows that this vanishes at the minimum of the potential. The scalar
potential is used to determine the vacuum, its energy, and any flat directions
it may have. Also, small deviations around it determine the
supersymmetry-breaking masses and couplings of the scalar fields. Finally,
derivatives of the K\"ahler function determine the supersymmetry-breaking
masses of the (non-chiral) fermions, and derivatives of the gauge kinetic
function determine the supersymmetry-breaking gaugino masses.

Spontaneous breakdown of supergravity induces a mass for the gravitino
\begin{equation}
m_{3/2}=\VEV{e^{G/2}}=\VEV{e^{K/2}\,|W|}\ .
\label{m3/2def}
\end{equation}
This relation shows that supersymmetry breaking can only occur if
$\VEV{W}\not=0$. All soft-supersymmetry-breaking parameters are proportional
to $m_{3/2}$, with typically ${\cal O}(1)$ coefficients of proportionality.
Therefore, $m_{3/2}$ values are expected to be not much higher than the
electroweak scale. Moreover, restoring the dimensions in Eq.~(\ref{m3/2def})
one sees that the right-hand-side has units of $10^{18}\GeV$ and therefore
a strong suppression of $e^{K/2}$ or $\VEV{W}$ is typically necessary. Two
scenarios for $\VEV{W}\not=0$ have received the most attention in the
literature: gaugino condensation in the hidden sector \cite{GauginoConden},
giving $W\sim e^{-b\pi^2/g^2}$; and string tree-level breaking via
coordinate-dependent compactifications \cite{Coordinate}, giving $W\sim1$.
\bigskip

No-scale supergravity is defined by three constraints on a supergravity model:
\begin{itemize}
\item The vacuum energy vanishes ($V_0=\VEV{V}=0$) by suitable choice of the
K\"ahler function ($G$) \cite{Cremmer}.
\item At the minimum of the potential there are flat directions (``moduli")
which leave the value of $m_{3/2}$ undetermined \cite{EKNI+II}.
\item The quantity ${\rm Str}\,{\cal M}^2$ should vanish at the minimum. This
constraint protects the potential from large one-loop corrections which would
otherwise force $m_{3/2}=0$ or $m_{3/2}=M_{Pl}$ \cite{EKNI+II}.
\end{itemize}
These three constraints impose severe restrictions on the possible $G$ and $f$
functions. Particularly non-trivial is the last one, \ie, ${\rm Str}\,{\cal
M}^2\equiv 2Qm^2_{3/2}$, with \cite{EKNI+II,FKZ}
\begin{equation}
Q=N-1-G^I(R_{I\bar J}-H_{I\bar J})G^{\bar J}\ ,
\label{Qdef}
\end{equation}
where $N$ is the total number of chiral superfields, and
\begin{eqnarray}
R_{I\bar J}&=&\partial_I\partial_{\bar J}\ln\det G_{M\bar N}\ ,\label{Rdef}\\
H_{I\bar J}&=&\partial_I\partial_{\bar J}\ln\det {\rm Re\,}(f_{ab})\
.\label{Hdef}
\end{eqnarray}

If the above three conditions are satisfied, the low-energy theory, obtained by
renormalization-group evolution from the Planck scale down to the electroweak
scale, will be undetermined to the extent that $m_{3/2}$ is undetermined, as it
depends on the undetermined moduli vacuum expectation values (VEVs). The
low-energy one-loop effective potential ($V_{\rm eff}$) then depends on the
usual Higgs fields, as well as the moduli fields. The {\em no-scale mechanism}
\cite{Lahanas} consists of minimizing this potential with respect to all these
fields, thus determining the Higgs {\em and} moduli vacuum expectation values.
The thusly determined moduli VEVs then determine $m_{3/2}$, and therefore all
of the supersymmetry breaking masses.

The no-scale mechanism has an additional unsuspected consequence: it may solve
the strong CP problem \cite{thetaQCD}. Indeed, the equally undetermined
imaginary parts of the moduli fields leave the $\theta_{\rm QCD}$ parameter
undetermined, \ie, the potential in the imaginary directions is also
flat.\footnote{This is certainly the case at tree-level in the K\"ahler
potential and for moduli-independent Yukawa couplings. In free-fermionic
models, moduli dependence of the superpotential does not arise until the
quartic order \cite{modinv}.} According to the usual argument, non-perturbative
QCD dynamics at low energies determines $\theta_{\rm QCD}=0$, which in our
language corresponds to lifting the imaginary flat directions, giving zero VEVs
to the corresponding fields.

In practice, this procedure is subtle and complicated by the existence of a
remnant vacuum energy term at high energies. This field-independent term
($Cm_{3/2}^4$) needs to be added to the low-energy effective potential to
ensure its renormalization-scale independence \cite{aspects,KPZ,No-scale}, \ie,
\begin{equation}
V_{\rm eff}=V_{\rm tree}+{1\over64\pi^2}{\rm Str}\,{\cal M}^4\left(\ln{{\cal
M}^2\over Q^2}-{3\over2}\right)-Cm_{3/2}^4\ ,
\label{V1def}
\end{equation}
where $V_{\rm tree}$ is the tree-level Higgs potential. It is worth mentioning
that just as the usual radiative breaking mechanism (\ie, minimization of
$V_{\rm eff}$ with respect to the Higgs fields) does not always work, the
no-scale mechanism may also not work. This happens when the effective potential
does not have a good minimum in the moduli directions. In the case of a single
modulus field (as we discuss below), it can be shown that a necessary condition
for a good minimum is ${\rm Str}\,{\cal M}^4>0$ \cite{No-scale}. If the
explicit $m_{3/2}$-dependent contribution to ${\rm Str}\,{\cal M}^4$ is
negative, then \cite{No-scale}
\begin{equation}
{\rm Str}\,{\cal M}^4>0\ \Longrightarrow\ {m_{3/2}\over m_{\tilde q}}\lsim{\cal
O}(1)\ ,
\label{Str>0}
\end{equation}
which imposes restrictions on the allowed low-energy parameter space.

\section{Fermionic string models}
\label{Fermionic}
The discussion in the previous section can be applied to any supergravity or
superstring model. However, the requirement of ${\rm Str}\,{\cal M}^2=0$ can
only be investigated if the full spectrum of the model is known, as in string
models. We are interested in exploring the three postulates of string no-scale
supergravity in the context of string models built within the free-fermionic
formulation \cite{FFF}. Our motivation for such choice is that fairly realistic
models already exist in this construction \cite{revamped,others,search,Moscow},
and we would like to know whether this class of models satisfies the
postulates, or what constraints may need to be imposed on model-building so
that these postulates are satisfied.

\subsection{Generalities}
\label{Generalities}
All of the level-one free-fermionic models built to date have been based on the
simplest supersymmetry-generating basis vector $S$. This choice is not unique,
but should suffice for our present purposes of investigating the viability of
no-scale supergravity in free-fermionic models. In this class of models, all
states in the spectrum fall into three {\em sets}, depending on the quantum
numbers they carry with respect to some internal symmetries of the
two-dimensional world-sheet theory \cite{KLN}. The spectrum further divides
itself into two sectors: untwisted and twisted.  More specifically, all matter
fields carry charges under three world-sheet $U(1)$ currents. The sum of these
three currents provides the additional current which extends the manifest $N=1$
world-sheet supersymmetry to $N=2$ world-sheet supersymmetry \cite{KLN}, as
required for $N=1$ space-time supersymmetry to exist. The scalar components of
untwisted (or Neveu-Schwarz) matter superfields carry one of three possible
types of charges, whereas the twisted matter superfields carry two of them,
\ie,
\begin{equation}
\begin{tabular}{lcc}
&Untwisted&Twisted\\
First set&$\{1,0,0\}$&$\{0,\half,\half\}$\\
Second set&$\{0,1,0\}$&$\{\half,0,\half\}$\\
Third set&$\{0,0,1\}$&$\{\half,\half,0\}$
\end{tabular}
\label{charges}
\end{equation}
The charges of their fermionic partners are obtained by a uniform shift of
$\{-\half,-\half,-\half\}$. From these charge assignments it follows
immediately what kind of cubic superpotential couplings are allowed in a model.
Indeed, the sum of the charges of the three fields in question must be
$\{1,1,1\}$; charge conservation follows from the two required shifts by
$\{-\half,-\half,-\half\}$, since a cubic coupling contains two fermionic
fields and one scalar field. Thus, one gets five types of cubic couplings
\begin{equation}
\begin{tabular}{ccc}
$U^{(1)}\,U^{(2)}\,U^{(3)}$&$U^{(1)}\,T^{(1)}\,T^{(1)}$&
$T^{(1)}\,T^{(2)}\,T^{(3)}$\\
&$U^{(2)}\,T^{(2)}\,T^{(2)}$&\\
&$U^{(3)}\,T^{(3)}\,T^{(3)}$&\\
\end{tabular}
\label{couplings}
\end{equation}
where $U^{(I)}$ [$T^{(I)}$] denotes a generic untwisted (twisted) field
belonging to the $I$-th set.

Several features of the K\"ahler function for the untwisted sector of such
models had been known for some time \cite{oldUT}, and have been recently
further clarified, extended, and applied to specific models in
Ref.~\cite{LNY94}. The corresponding contributions from the twisted
sector have been known for some time in simple models \cite{oldTS}, and have
been only recently calculated in realistic models \cite{LNY95}.
Abstracting all that is known about free-fermionic models, we write
\begin{equation}
G=-\ln(S+\bar S)+\sum_{I=1,2,3} K_{(I)}+K_{\rm TS}+\ln|W|^2\ ,
\label{GFFF}
\end{equation}
with
\begin{equation}
K_{(I)}=-\ln\left[1-\sum_i^{n_I}
\alpha_i\bar\alpha_i+\coeff{1}{4}\left(\sum_i^{n_I}\alpha_i^2\right)
\left(\sum_i^{n_I}\bar\alpha_i^2\right)
\right]\ ,
\label{Kdef}
\end{equation}
where $n_I$ represents the number of untwisted fields in set $I$, and
set-indices $I=1,2,3$ on the $\alpha_i$ (\ie, $\alpha^{(I)}_i$) are understood.
(The twisted sector contribution $K_{\rm TS}$ will be addressed below.) The
K\"ahler function in Eqs.~(\ref{GFFF}),(\ref{Kdef}) is written in the ``string
basis". For purposes of low-energy effective supergravity analyses, it is
more convenient to make suitable field redefinitions of the $\alpha_i$ to
exhibit the moduli fields that may be present in the spectrum, \ie, to go from
the ``string basis" to the ``supergravity basis". In the class of fermionic
models which we consider here, three possibilities for the untwisted moduli
space of any of the sets were identified in Ref.~\cite{LNY94}:
\begin{description}
\item (i) $[SU(1,1)/U(1)]^2$,
with two moduli fields denoted by $\tau_1,\tau_2$ (a ``$\tau_1,\tau_2$ set").
\footnote{In Ref.~\cite{LNY94}, $\tau_1,\tau_2$ were denoted by $T,U$. Such
notation would cause confusion here.}
\item (ii) $SU(1,1)/U(1)$ with one modulus field denoted by $\tau$ (a ``$\tau$
set").
\item (iii) No moduli at all (an ``$\alpha$ set").
\end{description}
Whichever of the three possibilities may be realized for a given set depends on
the choice of basis and GSO projections of the fermionic model. In what
follows, if a set has any modular symmetry at all, we perform a field
redefinition of the fields in that set which leaves the K\"ahler function ($G$)
unchanged. (Details of these manipulations are given in Appendix A.)
The redefined $K_{(I)}$ are given by:
\begin{itemize}
\item $\tau_1,\tau_2$ set:
\begin{equation}
K=-\ln\left[(\tau_1+\bar\tau_1)(\tau_2+\bar\tau_2)-\sum_i^{n_\phi}(\phi_i+\bar\phi_i)^2\right]\ ,
\label{TUdef}
\end{equation}
where $n_\phi=n_I-2$, since two of the $\alpha_i$ are transformed into the
moduli $\tau_1,\tau_2$. The scalar fields parametrize the K\"ahler manifold
$SO(2,2+n_\phi)/SO(2)\times SO(2+n_\phi)$, which has as a subspace the moduli
space $SO(2,2)/SO(2)\times SO(2)\approx[SU(1,1)/U(1)]^2$.
\item $\tau$ set:
\begin{equation}
K=-\ln\left[(\tau+\bar \tau)^2-\sum_i^{n_\psi}(\psi_i+\bar\psi_i)^2\right]\ ,
\label{taudef}
\end{equation}
where $n_\psi=n_I-1$, since one of the $\alpha_i$ is transformed into the
modulus $\tau$.  The scalar fields parametrize the K\"ahler manifold
$SO(2,1+n_\psi)/SO(2)\times SO(1+n_\psi)$,\footnote{We note that in the old
no-scale models \cite{EKNI+II}, the K\"ahler potential was assumed to be of the
form $K\propto \ln(T+\bar T - \sum_i C_i \bar C_i)$, which yields the metric of
the space $SU(1,n_C+1)/U(1)\times SU(n_C+1)$. The old no-scale ansatz and the
string free-fermionic $\tau$-set result agree only for $n_\psi=n_C=0$.} which
has as a subspace the moduli space $SO(2,1)/SO(2)\approx SU(1,1)/U(1)$.
\end{itemize}

The modular symmetries exhibited above can be extended from the K\"ahler
potential ($K$) to the whole K\"ahler function ($G=K+\ln|W|^2$) if the matter
fields have suitable transformation properties under the modular symmetries
\cite{LNY94}. In fact, the superpotential should transform as a modular form
of weight $-1$. For instance, if the modulus field in question belongs to the
first untwisted set, then the modular weights of the matter fields are the
negative of the first world-sheet $U(1)$ charge in Eq.~(\ref{charges}), \ie,
\begin{equation}
\begin{tabular}{crccr}
$U^{(1)}$& $-1$ &\qquad\qquad &$T^{(1)}$& $0$\\
$U^{(2)}$& $0$ & &$T^{(2)}$& $-\coeff{1}{2}$\\
$U^{(3)}$& $0$ & &$T^{(3)}$& $-\coeff{1}{2}$\\
\end{tabular}
\label{ModularWeights}
\end{equation}
With this modular weight assignment one can verify that all cubic
superpotential couplings in Eq.~(\ref{couplings}) have modular weight $-1$. A
similar argument holds for moduli in the other sets. When quartic or
higher-order superpotential couplings are considered, the resulting modular
weight imbalance has to be compensated by the insertion of suitable powers
of the Dedekind eta function \cite{modinv}.

It is important to reiterate that modular
symmetries inferred from the K\"ahler potential must be respected by the whole
K\"ahler function. If presumed moduli fields appear in the cubic
superpotential, then the symmetry is explicitly broken and the presumed moduli
are to be discarded. This phenomenon is quite common in free-fermionic models
and has the beneficial effect of reducing the number of moduli in the model.
Therefore, in what follows, when discussing the number and type of moduli and
their impact on the vacuum energy and other properties of the models, we refer
to fields which have been indentified as untwisted moduli {\em and} that have
no superpotential couplings.

The twisted sector contribution to the K\"ahler potential
($K_{\rm TS}$) has been obtained some time ago for simple models with a
specific type and number of twisted sectors \cite{oldTS}. It turns out that the
result found in Ref.~\cite{oldTS} in fact applies to realistic free-fermionic
models
with a large number of twisted sectors \cite{LNY95} (see Appendix~\ref{AppB}
for details). The result is
\begin{equation}
K_{\rm TS}=
 \sum_i^{n_{T1}} \beta^{(1)}_i\bar\beta^{(1)}_i\ e^{{1\over2}[K_{(2)}+K_{(3)}]}
+\sum_i^{n_{T2}} \beta^{(2)}_i\bar\beta^{(2)}_i\ e^{{1\over2}[K_{(1)}+K_{(3)}]}
+\sum_i^{n_{T3}} \beta^{(3)}_i\bar\beta^{(3)}_i\ e^{{1\over2}[K_{(1)}+K_{(2)}]}
\label{KTS}
\end{equation}
where the $\beta^{(I)}_i$ are twisted sector fields that belong to the $I$-th
set, $n_{T1,T2,T3}$ are the numbers of these fields, and $K_{(1,2,3)}$ are
given in Eq.~(\ref{Kdef}). It is important to realize that this result is only
valid to lowest order in the twisted matter fields.

\subsection{Computation of $V_0$}
The computation of the scalar potential in Eq.~(\ref{Vdef}) requires the
knowledge of $G^{I\bar J}$, \ie, the transpose of the inverse of the matrix of
second derivatives of the K\"ahler potential. The computation is simplified
by that fact that the matrix $G_{I\bar J}=K_{I\bar J}$ possesses a
block-diagonal form once the twisted sector matter fields are set at their
zero vacuum expectation values. Indeed, schematically we have
\begin{equation}
K_{I\bar J}\sim
\bordermatrix{&U^{(1)}&U^{(2)}&U^{(3)}&T^{(1)}&T^{(2)}&T^{(3)}\cr
U^{(1)}&X_{U^{(1)}}&[T^{(3)}]^2&[T^{(2)}]^2&0&T^{(2)}&T^{(3)}\cr
U^{(2)}&[T^{(3)}]^2&X_{U^{(2)}}&[T^{(1)}]^2&T^{(1)}&0&T^{(3)}\cr
U^{(3)}&[T^{(2)}]^2&[T^{(1)}]^2&X_{U^{(3)}}&T^{(1)}&T^{(2)}&0\cr
T^{(1)}&0&T^{(1)}&T^{(1)}&X_{T^{(1)}}&0&0\cr
T^{(2)}&T^{(2)}&0&T^{(2)}&0&X_{T^{(2)}}&0\cr
T^{(3)}&T^{(3)}&T^{(3)}&0&0&0&X_{T^{(3)}}\cr}\ .
\label{schematic}
\end{equation}
At this point of the calculation all derivatives have been taken and we can
set $\VEV{T^{(1,2,3)}}=0$, which reveals the block-diagonal structure with
\begin{equation}
K^{I\bar J}={\rm diag}\,\{
X^{-1}_{U^{(1)}},X^{-1}_{U^{(2)}},X^{-1}_{U^{(3)}},
X^{-1}_{T^{(1)}},X^{-1}_{T^{(2)}},X^{-1}_{T^{(3)}}\}\ .
\label{inv-schematic}
\end{equation}
(There is an additional contribution to $K^{I\bar J}$ from the dilaton field.)
For each of these blocks one can compute $G^IG_I=G^{I\bar J}G_{\bar J}G_I$.
This calculation depends on the superpotential since
$G_I=K_I+\partial_I \ln W$, \ie,
\begin{equation}
G^IG_I=K^IK_I+K^I\partial_I\ln W+K^{\bar I}\partial_{\bar I}\ln\overline W
+K^{I\bar J}\partial_I\ln W\partial_{\bar J}\ln\overline W\ .
\label{GIGI}
\end{equation}
With the help of Mathematica \cite{Mathematica} we obtain for the untwisted
fields
\begin{eqnarray}
&{\rm Dilaton:}\qquad &\left[K^IK_I\right]^{(S)}=1\ .\label{KKS}\\
&\tau_1,\tau_2\ {\rm set:}\qquad
&\left[K^IK_I\right]^{(\tau_1,\tau_2)}=2\quad(\forall\ n_\phi)\
.\label{KKTU}\\
&{\rm \tau\ set:}\qquad  &\left[K^IK_I\right]^{(\tau)}=2\quad(\forall\ n_\psi)\
.\label{KKtau}\\
&{\rm \alpha\ set:}\qquad
&\left[K^IK_I\right]^{(\alpha)}=\sum_i^{n_I}\alpha_i\bar\alpha_i\ .
\label{KKalpha}
\end{eqnarray}
We do not show the corresponding results for the twisted fields since
$\VEV{K^IK_I}=0$ in this case (\ie, $\VEV{K_\beta}\propto\VEV{\bar\beta}=0$).
The scalar potential then becomes
\begin{equation}
V=e^G\left[ \left\{\begin{array}{cc}1,&\lambda_f=1\\ (S+\bar
S)^2|G_S|^2,&\lambda_f=0\end{array}\right\}+2n_{\rm\tau_1\tau_2}
+2n_\tau+n_\alpha\sum_i^{n_I}\alpha_i\bar\alpha_i-3+F(\beta,\partial\ln W)
\right]\ ,
\label{V}
\end{equation}
where the sum of the three kinds of sets
$n_{\rm\tau_1\tau_2}+n_\tau+n_\alpha=3$ is
fixed. The term $n_\alpha\sum_i\alpha_i\bar\alpha_i$ is meant to represent
however many $\alpha$-set contributions may exist in a given model. Also,
$\lambda_f=1$ indicates that $W$ does not depend on $S$, whereas $\lambda_f=0$
indicates that $W$ does depend on $S$ in which case $G_S=-1/(S+\bar
S)+(\partial_S W)/W$. The contributions which depend on $\partial_I\ln W$,
$\partial_{\bar I}\ln\overline W$, or the twisted fields are collectively
denoted by $F(\beta,\partial\ln W)$. The minimum of the potential is given by
\begin{equation}
V_0=e^G\,[\lambda_f+2n_{\rm \tau_1\tau_2}+2n_\tau-3]\ ,
\label{V0}
\end{equation}
with $\VEV{\alpha}=\VEV{\beta}=0$, $\VEV{\partial_I\ln W}=0$,\footnote{Since
$F(\beta,\partial\ln W)$ is generally a complicated expression, it may be
possible to find special minima for particular non-zero values of
$\VEV{\partial_I \ln W}$. In the case of the old no-scale models,
$F(\beta,\partial\ln W)=\sum_i |\partial_{C_i}\ln W|^2>0$ \cite{EKNI+II}, and
therefore $\VEV{\partial_{C_i}\ln W}=0$ is required.} and {\em if}
$\lambda_f=0$ also \hbox{$\VEV{-1/(S+\bar S)+(\partial_S W)/W}=0$.}

In view of Eq.~(\ref{V0}), there are only two choices for the untwisted modular
symmetry which are consistent with $V_0=0$, namely
\begin{equation}
\begin{tabular}{cccc}
$\lambda_f$&$n_{\rm \tau_1\tau_2}$&$n_\tau$&$n_\alpha$\\
1&1&0&2\\
1&0&1&2
\end{tabular}\qquad
\begin{tabular}{c}
Moduli\\
$S,\tau_1,\tau_2$\\
$S,\tau$
\end{tabular}
\label{V0=0}
\end{equation}
Note that the dilaton is required to be a modulus field, and that only one set
contributes moduli fields. If these requirements are satisfied,
we would obtain a model with zero vacuum energy and flat directions,
thus satisfying the first two no-scale supergravity postulates. Moreover,
the gravitino mass is given by (see Eq.~(\ref{m3/2def}))
\begin{equation}
m^2_{3/2}=\left\{\begin{array}{c}
{\VEV{|W|^2}\over
\VEV{(S+\bar
S)\,[(\tau_1+\bar\tau_1)(\tau_2+\bar\tau_2)-\sum_i(\phi_i+\bar\phi_i)^2]}}\\
{\VEV{|W|^2}\over
\VEV{(S+\bar S)\,[(\tau+\bar\tau)^2-\sum_i(\psi_i+\bar\psi_i)^2]}}
\end{array}\right.
\label{m3/2}
\end{equation}
for each of the two cases in Eq.~(\ref{V0=0}), and is undetermined as
anticipated. In these equations the values of $\VEV{\phi_i,\psi_i}$ are
determined by the flatness conditions $\VEV{\partial_I\ln W}=0$ (typically
$\VEV{\phi_i}=\VEV{\psi_i}=0$).

\subsection{Computation of $Q$}
\label{computationofQ}
We now compute $Q$ using the formula in Eq.~(\ref{Qdef}). The basic quantity
to be computed is the determinant of $G_{M\bar N}=K_{M\bar N}$. Since we
know the untwisted sector contribution to $K$ exactly, whereas we only
have a first-order approximation to the twisted sector contribution (see
Eq.~(\ref{KTS})), we address the untwisted sector first. As in the
computation of $G^IG_I$ above, the $K_{M\bar N}$ matrix is block-diagonal
(three untwisted
sets and the dilaton) and (with the help
of Mathematica) we obtain
\begin{eqnarray}
&{\rm Dilaton:}\qquad &\left[\det G_{M\bar N}\right]^{(S)}=(S+\bar S)^{-2}
\ .\label{detS}\\
&\tau_1,\tau_2\ {\rm set:}\qquad  &\left[\det G_{M\bar
N}\right]^{(\tau_1,\tau_2)}=2^{n_\phi}
\left[(\tau_1+\bar\tau_1)(\tau_2+\bar\tau_2)-
\sum_i^{n_\phi}(\phi_i+\bar\phi_i)^2\right]^{-n_\phi-2}\label{detTU}\\
&{\rm \tau\ set:}\qquad  &\left[\det G_{M\bar N}\right]^{(\tau)}=
2^{n_\psi+1}\left[(\tau+\bar \tau)^2-
\sum_i^{n_\psi}(\psi_i+\bar\psi_i)^2\right]^{-n_\psi-1}\label{dettau}\\
&{\rm \alpha\ set:}\qquad  &\left[\det G_{M\bar N}\right]^{(\alpha)}=
\left[1-\sum_i^{n_I}
\alpha_i\bar\alpha_i+\coeff{1}{4}\left(\sum_i^{n_I}\alpha_i^2\right)
\left(\sum_i^{n_I}\bar\alpha_i^2\right)\right]^{-n_I}\label{detalpha}
\end{eqnarray}
{}From these results and Eqs.~(\ref{GFFF},\ref{Kdef},\ref{TUdef},\ref{taudef})
we see that $R_{I\bar J}=\partial_I\partial_{\bar J}\ln\det G_{M\bar N}$ for
each block is just a multiple of $G_{I\bar J}$ for that block, \ie,
\begin{eqnarray}
&{\rm Dilaton:}\qquad &\left[R_{I\bar J}\right]^{(S)}=2 G_{S\bar S}^{(S)}
\ .\label{RS}\\
&\tau_1,\tau_2\ {\rm set:}\qquad  &\left[R_{I\bar
J}\right]^{(\tau_1,\tau_2)}=(n_\phi+2)G_{I\bar
J}^{(\tau_1,\tau_2)}\label{RTU}\\
&{\rm \tau\ set:}\qquad  &\left[R_{I\bar J}\right]^{(\tau)}=(n_\psi+1)G_{I\bar
J}^{(\tau)}\label{Rtau}\\
&{\rm \alpha\ set:}\qquad  &\left[R_{I\bar J}\right]^{(\alpha)}=n_I G_{I\bar
J}^{(\alpha)}\label{Ralpha}
\end{eqnarray}
With the above observation, the quantity that appears in $Q$ can be readily
obtained: for each block $G^IR_{I\bar J}G^{\bar J}\propto G^IG_{I\bar J}G^{\bar
J}=G^IG_I$, and these quantities have been given in
Eqs.~(\ref{KKS})--(\ref{KKalpha}) (at the minimum $\VEV{G^IG_I}=\VEV{K^IK_I}$).

The computation of $Q$ also involves evaluating $H_{I\bar
J}=\partial_I\partial_{\bar J}\ln\det {\rm Re\,}(f_{ab})$. In string models,
$f_{ab}$ receives tree-level and one-loop contributions only \cite{f1loop}.
Writing $f_{ab}=\delta_{ab}S+f^{\rm 1-loop}$ (suitable for level-one Kac-Moody
constructions), and neglecting the one-loop contribution, we obtain
$\det {\rm Re\,}(f_{ab})=[{1\over2}(S+\bar S)]^{d_f}$, where $d_f$ is the
dimension of the gauge group ($d_f\gg1$). Also, $H_{S\bar S}=-d_f G_{S\bar S}$
and the contribution to $Q$ is $G^IH_{I\bar J}G^{\bar J}=-d_f G^SG_S=-\lambda_f
d_f$ (at the minimum).

The total contribution to $Q$ is then
\begin{eqnarray}
Q&=&\left[1+(2+n_\phi)n_{\rm \tau_1\tau_2}+(1+n_\psi)n_\tau+n_I n_\alpha\right]
 - 1\nonumber\\
&&-\left[\{2\lambda_f+2(n_\phi+2)n_{\rm \tau_1\tau_2}
+2(n_\psi+1)n_\tau\}+\lambda_f d_f\right]\nonumber\\
&&+Q_{\rm TS}\ ,
\label{Qresult}
\end{eqnarray}
where the terms are displayed in correspondence with those in Eq.~(\ref{Qdef}),
and $Q_{\rm TS}$ is the twisted sector contribution to $Q$. For the two cases
in Eq.~(\ref{V0=0}), which give $V_0=0$, we obtain
\begin{equation}
Q=\left\{\begin{array}{c} 2n_I-n_\phi-4-d_f+Q_{\rm TS}\\
2n_I-n_\psi-3-d_f+Q_{\rm TS}\end{array}\right.\ ,
\label{QV0=0}
\end{equation}
where ``$2n_I$" is meant to represent the sum of the untwisted fields in the
two sets which do not contain moduli.

Now let us address the twisted sector contribution to $Q$ (\ie, $Q_{\rm TS}$).
Of the two cases in Eq.~(\ref{V0=0}), consistent with $V_0=0$, the second case
corresponds to a specific string model which will be discussed in
Sec.~\ref{Possible} below; we focus on this case in what follows. The complete
K\"ahler potential in this case is (see Appendix~\ref{AppB} for details)
\begin{eqnarray}
K&=&-\ln(S+\bar S)
-\ln\left[(\tau+\bar \tau)^2-\sum_i^{n_\psi}(\psi_i+\bar\psi_i)^2\right]
+\sum_i^{n_{U2}}\alpha^{(2)}_i\bar\alpha^{(2)}_i
+\sum_i^{n_{U3}}\alpha^{(3)}_i\bar\alpha^{(3)}_i\nonumber\\
+&&\!\!\!\!\!\!\sum_i^{n_{T1}}\beta^{(1)}_i\bar\beta^{(1)}_i
+{1\over\left[(\tau+\bar
\tau)^2-\sum_i^{n_\psi}(\psi_i+\bar\psi_i)^2\right]^{1/2}}
\left(\sum_i^{n_{T2}}\beta^{(2)}_i\bar\beta^{(2)}_i
+\sum_i^{n_{T3}}\beta^{(3)}_i\bar\beta^{(3)}_i\right)
\label{completeK}
\end{eqnarray}
which is valid to first order in the twisted fields $\beta_i^{(1,2,3)}$,
and the untwisted fields $\alpha_i^{(2,3)}$, and corresponds
to $2n_I=n_{U2}+n_{U3}$ in Eq.~(\ref{QV0=0}). The contribution to $Q_{\rm TS}$
from the first twisted set comes only from the $n_{T1}$ contribution to the
total number of chiral superfields ($N$) in Eq.~(\ref{Qdef}), since
$K_{\beta^{(1)}\bar\beta^{(1)}}={\bf1}$ has unit determinant and thus
$R_{\beta^{(1)}\bar\beta^{(1)}}=0$. The second and third twisted sets
contribute to $N$ ($n_{T2},n_{T3}$), and in principle get tangled up with the
fields in the first untwisted set ($\tau,\psi_i$). However, since
$\VEV{G_{\beta^{(2,3)}}}=0$, we only need to worry about their possible
additional contributions to $[R_{I\bar J}]^{(\tau)}$ in Eq.~(\ref{Rtau}). This
means that in this enlarged determinant (involving
$\tau,\psi_i,\beta^{(2)}_i,\beta^{(3)}_i$) we can set $\beta^{(2,3)}_i=0$ after
calculating the derivatives but before calculating the determinant. The
enlarged determinant is then the one given in Eq.~(\ref{dettau}) times the
factor $\left[(\tau+\bar
\tau)^2-\sum_i^{n_\psi}(\psi_i+\bar\psi_i)^2\right]^{-(n_{T2}+n_{T3})/2}$.
That is, the exponent in Eq.~(\ref{dettau}) and the coefficient in
Eq.~(\ref{Rtau}) receive a further contribution of $-(n_{T2}+n_{T3})/2$.
The result for $Q$ is then
\begin{eqnarray}
Q&=&[1+(1+n_\psi)+n_{U2}+n_{U3}+n_{T1}+n_{T2}+n_{T3}]-1\nonumber\\
&&-[\{2+2(n_\psi+1+(n_{T2}+n_{T3})/2)\}+d_f]\nonumber\\
&=&n_{U2}+n_{U3}+n_{T1}-n_\psi-d_f-3\ .
\label{Qformula}
\end{eqnarray}

What are the prospects for obtaining $Q=0$? In typical models one observes that
$n_{U2}\sim n_{U3}\sim n_\psi\sim{\cal O}(10)\ll n_{T1}$. On the other hand
$d_f>d_{\rm SM}=12$, although in realistic models we expect a number
significantly exceeding this lower bound since, \eg, the hidden sector gauge
group needs to be large enough for supersymmetry breaking via gaugino
condensation to occur at a sufficiently high scale; a typical value would be
$d_f\sim{\cal O}(50-100)$. Thus, it is not inconceivable that models can be
found where the various contributions to $Q$ cancel each other out. In
Sec.~\ref{Possible} we exhibit a model where this cancellation is almost
perfect.

\section{Soft-supersymmetry-breaking parameters}
\label{Soft}
With the knowledge of the K\"ahler function, the scalar potential, and the
gauge kinetic function one can compute the usual soft-supersymmetry-breaking
parameters on which the low-energy model predictions depend so crucially.

\subsection{Goldstino composition}
Before we proceed with these calculations, it is instructive to determine the
field dependence of the goldstino field, which has received considerable
attention in ``model-independent" approaches to this problem.
The goldstino, which is eaten by the gravitino upon spontaneous breaking of
supergravity, is given by
\begin{equation}
\widetilde\eta=\VEV{e^{G/2}G_I}\chi^I\ ,
\label{etadef}
\end{equation}
where $\chi^I$ are the fermionic partners of the scalar fields which appear
in the scalar potential (\ref{Vdef}). At the minimum of the scalar potential
we have $\VEV{G_{\alpha,\beta}}=0$. In computing the $G_I$ derivatives,
for present purposes it suffices to approximate the K\"ahler function in
Eqs.~(\ref{TUdef}),(\ref{taudef}) in the limit where the $\psi_i,\phi_i$ fields
have vevs much smaller than the moduli vevs. This gives
\begin{eqnarray}
&{\rm Dilaton:}\qquad &\VEV{G_I}^{(S)}=-\lambda_f(S+\bar S)^{-1}\label{G_IS}\\
&\tau_1,\tau_2\ {\rm set:}\qquad
&\VEV{G_I}^{(\tau_1,\tau_2)}\approx\{-(\tau_1+\bar\tau_1)^{-1},-(\tau_2
+\bar\tau_2)^{-1},\nonumber\\
&&\qquad\qquad\qquad\qquad 2(\phi_i+\bar\phi_i)[(\tau_1+\bar\tau_1)(\tau_2
+\bar\tau_2)]^{-1}\}\label{G_ITU}\\
&{\rm \tau\ set:}\qquad  &\VEV{G_I}^{(\tau)}\approx\{-2(\tau+\bar\tau)^{-1},
2(\psi_i+\bar\psi_i)(\tau+\bar\tau)^{-2}\}\label{G_Itau}
\end{eqnarray}
These results can be substituted back into Eq.~(\ref{etadef}) to obtain
$\widetilde\eta$. The final step is to express the $\chi^I$ fields in terms of
the properly normalized $\widehat\chi^I$ fields. This operation entails a
rescaling of the fields which will be discussed in detail in Sec.~\ref{Field}.
Let us just quote the results:
\begin{eqnarray}
\widehat S&=&{S\over\VEV{S+\bar S}}\ ,\quad \widehat
\tau_1={\tau_1\over\VEV{\tau_1+\bar\tau_1}}\
,\quad
\widehat \tau_2={\tau_2\over\VEV{\tau_2+\bar\tau_2}}\ ,\quad
\widehat\tau={\sqrt{2}\,\tau\over\VEV{\tau+\bar\tau}}\ ,
\label{redefinitions1}
\\
\widehat\phi_i&=&{\sqrt{2}\,\phi_i\over\VEV{(\tau_1+\bar\tau_1)(\tau_2+\bar\tau_2)}^{1/2}}\
,\qquad
\widehat\psi_i={\sqrt{2}\,\psi_i\over\VEV{\tau+\bar\tau}}\ .
\label{redefinitions2}
\end{eqnarray}
The goldstino fields corresponding to the two cases in Eq.~(\ref{V0=0}) are
then
\begin{equation}
\widetilde\eta\propto\left\{
\begin{array}{l}
\widehat S+\widehat\tau_1+\widehat\tau_2
+\sqrt{2}\,\sum_i^{n_\phi}{\VEV{\phi_i+\bar\phi_i}\over\VEV{(\tau_1
+\bar\tau_1)(\tau_2+\bar\tau_2)}^{1/2}}\,(\widehat\phi_i+\widehat{\bar\phi_i})\\\widehat S+\sqrt{2}\,\widehat\tau+
\sqrt{2}\,\sum_i^{n_\psi}{\VEV{\psi_i+\bar\psi_i}\over\VEV{\tau+\bar\tau}}\,
(\widehat\psi_i+\widehat{\bar\psi_i})
\end{array}\right.
\longrightarrow
\widetilde\eta=\left\{
\begin{array}{l}
{1\over\sqrt{3}}(\widehat S+\widehat\tau_1+\widehat\tau_2)\\
{1\over\sqrt{3}}(\widehat S+\sqrt{2}\,\widehat\tau)
\end{array}\right.
\label{goldstino}
\end{equation}
where the second form holds in the limit $\VEV{\psi_i,\phi_i}\approx0$.
Thus, we get a goldstino field which contains substantial components of
both ``dilaton" and ``moduli". Note that, in principle light matter fields
also appear, although their contribution is highly suppressed, by a factor
$\VEV{\phi}/\VEV{\rm moduli}\sim{\cal O}(10^2/10^{18})$.

\subsection{Gaugino masses}
The properly normalized gaugino masses are obtained from the expression
\begin{equation}
M_a={e^{G/2}\over 2{\rm Re}\, f_a}\sum_I \partial_I f_a\  G^I,
\label{Madef}
\end{equation}
where the sum over $I$ runs over all matter fields which $f_a$ depends on.
For $f_a$ we use the following one-loop (although correct to all orders)
expression \cite{f1loop}
\begin{equation}
f_a=k_a S
-\coeff{1}{16\pi^2}B^{(\tau_1,\tau_2)}_a\ln|\eta(\tau_1)\eta(\tau_2)|^4\cdot
n_{\rm \tau_1\tau_2}
-\coeff{1}{16\pi^2}B^{(\tau)}_a\ln|\eta(\tau)|^4\cdot n_\tau\ ,
\label{fadef}
\end{equation}
where the level of the Kac-Moody algebra is one ($k_a=1$), and $\eta$ is the
Dedekind eta function. Also, $B_a$ is a quantity which depends on the massless
sector of the theory and their modular weights, as well as on the coefficient
$\delta_{\rm GS}$ which arises in the Green-Schwarz modular-anomaly
cancellation. For our present purposes, a detailed specification of $B_a$ is
not required. The derivatives $\partial_I f_a$
in the expression for $M_a$ are non-zero only for $S,\tau_1,\tau_2,\tau$. One
obtains,
\eg, $\partial_{\tau_1} f_a={1\over16\pi^3}B^{(\tau_1,\tau_2)}_a \widehat
G_2(\tau_1)$, where
$\widehat G_2(\tau_1)=G_2(\tau_1)-2\pi/(\tau_1+\bar\tau_1)$, and the Eisenstein
function $G_2$
is related to the Dedekind function via
$G_2(\tau_1)=-4\pi\partial_{\tau_1}\ln\eta(\tau_1)$
\cite{Cvetic}. We also note that $\widehat G_2(\tau_1)$ has zeroes at
$\tau_1=1,e^{i\pi/6}$.

The other ingredient in the expression for $M_a$ is $G^I=G^{I\bar J}G_{\bar
J}=K^{I\bar J}(K_{\bar J}+\partial_{\bar J}\ln\overline W)=
K^{I\bar J}K_{\bar J}=K^I$ at the minimum. This expression can be evaluated,
with the result
\begin{eqnarray}
&{\rm Dilaton:}\qquad &[K^I]^{(S)}=-(S+\bar S)\ ,\label{K^IS}\\
&\tau_1,\tau_2\ {\rm set:}\qquad
&[K^I]^{(\tau_1,\tau_2)}=\{-(\tau_1+\bar\tau_1),-(\tau_2+\bar
\tau_2),-(\phi_i+\bar\phi_i)\}\ ,\label{K^ITU}\\
&{\rm \tau\ set:}\qquad  &[K^I]^{(\tau)}=\{-(\tau+\bar\tau),
-(\psi_i+\bar\psi_i)\}\ ,\label{K^Itau}\\
&{\rm \alpha\ set:}\qquad  &[K^I]^{(\alpha)}=\{\alpha_i+{\cal O}(\alpha^2_i)\}\
.\label{K^Ialpha}
\end{eqnarray}
With these results we finally obtain (only $I=S,\tau_1,\tau_2,\tau$ are
relevant)
\begin{eqnarray}
M_a={m_{3/2}\over2{\rm Re}\, f_a}&&\hspace{-0.5cm}\Bigl\{-(S+\bar
S)\lambda_f\nonumber\\
&&-\coeff{1}{16\pi^2}B^{(\tau_1,\tau_2)}_a[(\tau_1+\bar\tau_1)\widehat
G_2(\tau_1)+(\tau_2+\bar\tau_2)\widehat
G_2(\tau_2)]\cdot n_{\rm \tau_1\tau_2}\nonumber\\
&&-\coeff{1}{16\pi^2}B^{(\tau)}_a(\tau+\bar\tau)\widehat G_2(\tau)\cdot
n_\tau\Bigr\}\ .
\label{Ma}
\end{eqnarray}
For the two cases in Eq.~(\ref{V0=0}) $\lambda_f=1$ and thus the tree-level
contribution to $M_a$ is non-zero and therefore dominant, giving nearly (up to
small one-loop corrections) {\em universal} gaugino masses, \ie,
\begin{equation}
M_a=m_{1/2}=m_{3/2}\ .
\label{m1/2}
\end{equation}

\subsection{Scalar masses}
The scalar masses are obtained by taking second derivatives of the scalar
potential.\footnote{One has to properly normalize the fields (\eg,
$\psi\to\widehat\psi$) to obtain the physical masses. The normalization factors
are given in
Eqs.~(\ref{redefinitions1},\ref{redefinitions2},\ref{normalizations1},
\ref{normalizations2},\ref{normalizations3}), and can be trivial (\ie, $1$) in
many instances.} These masses have two sources: ``supersymmetric" masses from
the superpotential, and supersymmetry-breaking masses from the K\"ahler
potential. For the untwisted fields, the latter can be deduced from the
expression for $V$ given in Eq.~(\ref{V}) (only the $K^IK_I$ term in
Eq.~(\ref{GIGI}) matters). We see that neither the $\phi_i$ nor the $\psi_i$
fields appear, thus
\begin{equation}
\widetilde m_{\widehat\phi_i}=\widetilde m_{\widehat\psi_i}=0\ ,
\label{mphipsi}
\end{equation}
where $\widetilde m_f$ represents the supersymmetry-breaking contribution to
the mass of the scalar field $f$. On the other hand, the $\alpha_i$ do appear
in $V$ and their mass is given by
\begin{equation}
\widetilde m_{\widehat\alpha_i}=m_{3/2}\ .
\label{malpha}
\end{equation}
In other words, untwisted sector fields in sets with moduli receive no
(tree-level) supersymmetry-breaking masses, whereas those in sets with no
moduli receive a universal mass equal to the gravitino mass. Obviously, it also
follows that the moduli are massless (including the dilaton since $\lambda_f=1$
is required).

Turning to the twisted field scalar masses, let us again consider the second
case in Eqs.~(\ref{V0=0},\ref{QV0=0}), with the K\"ahler potential given in
Eq.~(\ref{completeK}). For the fields in the first untwisted set
[$\beta^{(1)}_i$] it is clear that their (K\"ahler potential) masses are equal
to those of the second [$\alpha^{(2)}_i$] and third [$\alpha^{(3)}_i$]
untwisted set fields. Indeed, in this case
$K_{\beta^{(1)}\bar\beta^{(1)}}={\bf1}$ and
$K^IK_I=\sum_i^{n_{T1}}\beta^{(1)}_i\bar\beta^{(1)}_i$, and as expected
\begin{equation}
\widetilde m_{\widehat\beta^{(1)}_i}=m_{3/2}\ .
\label{mbeta1}
\end{equation}
The scalar masses of the second and third twisted set fields can be obtained by
considering the following portion of the K\"ahler potential in
Eq.~(\ref{completeK})
\begin{equation}
K_0(\tau,\bar\tau)+
K_1(\tau,\bar\tau)\sum_i^{n_\psi}(\psi_i+\bar\psi_i)^2
+K_2(\tau,\bar\tau)\left(\sum_i^{n_{T2}}\beta^{(2)}_i\bar\beta^{(2)}_i
+\sum_i^{n_{T3}}\beta^{(3)}_i\bar\beta^{(3)}_i\right)\ ,
\label{ApproxK}
\end{equation}
with
\begin{equation}
K_0=-\ln(\tau+\bar \tau)^2\,,\qquad K_1={1\over(\tau+\bar\tau)^2}\,,\qquad
K_2={1\over\tau+\bar \tau}\ ,
\label{K0K1K2}
\end{equation}
where we have performed an expansion to first order in both $\psi_i$ and
$\beta_i^{(2,3)}$. The approximate expression in Eq.~(\ref{ApproxK}) is
sufficient to compute the scalar masses. After some algebra we obtain for this
subset of the fields (\ie, $\tau,\psi_i,\beta^{(2)}_i,\beta^{(3)}_i$)
\begin{eqnarray}
K^IK_I&=&{K_{0\tau}K_{0\bar\tau}\over K_{0\tau\bar\tau}}\nonumber\\
&&+2K_1\left[1
-{K_{0\tau}K_{1\bar\tau}+K_{0\bar\tau}K_{1\tau}\over 2K_1K_{0\tau\bar\tau}}
-{K_{0\tau}K_{0\bar\tau}\over K_{0\tau\bar\tau}}
{K_1K_{1\tau\bar\tau}-2K_{1\tau}K_{1\bar\tau}\over 2(K_1)^2K_{0\tau\bar\tau}}
\right]\sum_i^{n_\psi}(\psi_i+\bar\psi_i)^2\nonumber\\
&&+K_2\left[1-{K_{0\tau}K_{0\bar\tau}\over K_{0\tau\bar\tau}}\,{(\ln
K_2)_{\tau\bar\tau}\over K_{0\tau\bar\tau}}\right]
\left(\sum_i^{n_{T2}}\beta^{(2)}_i\bar\beta^{(2)}_i
+\sum_i^{n_{T3}}\beta^{(3)}_i\bar\beta^{(3)}_i\right)\ .
\label{KIKIapprox}
\end{eqnarray}
That is
\begin{eqnarray}
\widetilde m^2_{\widehat\psi_i}&=&1
-{K_{0\tau}K_{1\bar\tau}+K_{0\bar\tau}K_{1\tau}\over 2K_1K_{0\tau\bar\tau}}
-{K_{0\tau}K_{0\bar\tau}\over K_{0\tau\bar\tau}}
{K_1K_{1\tau\bar\tau}-2K_{1\tau}K_{1\bar\tau}\over 2(K_1)^2K_{0\tau\bar\tau}}
\ ,
\label{psimass}
\\
\widetilde m^2_{\widehat\beta^{(2)}_i}&=&\widetilde m^2_{\widehat\beta^{(3)}_i}
=1-{K_{0\tau}K_{0\bar\tau}\over K_{0\tau\bar\tau}}\,{(\ln
K_2)_{\tau\bar\tau}\over K_{0\tau\bar\tau}}
\ ,
\label{betamass}
\end{eqnarray}
where we have properly normalized the $\psi_i$ and $\beta^{(2,3)}_i$ fields
by absorbing the overall factors (see
Eqs.~(\ref{normalizations1},\ref{normalizations3})).
For the choices of $K_0,K_1,K_2$ in Eq.~(\ref{K0K1K2}), the first term in
Eq.~(\ref{KIKIapprox}) is $K_{0\tau}K_{0\bar\tau}/K_{0\tau\bar\tau}=2$, as
expected from the vacuum energy calculation above. It also follows that
$\widetilde m^2_{\widehat\psi_i}=0$ (confirming the result in
Eq.~(\ref{mphipsi})) and the new result
\begin{equation}
\widetilde m_{\widehat\beta^{(2)}_i}=\widetilde m_{\widehat\beta^{(3)}_i}=0\ .
\label{mbeta2beta3}
\end{equation}
The expression for the $\beta^{(2,3)}_i$ masses in Eq.~(\ref{betamass}) agrees
with that given in Ref.~\cite{BIM};  the expression for the $\psi_i$ masses is
new.

To summarize, for the model with K\"ahler potential given in
Eq.~(\ref{completeK}), the scalar masses of all fields are the following
multiples of $m_{3/2}$
\begin{equation}
\begin{tabular}{ccrcccr}
$U^{(1)}\,$:&$\psi_i$& $0$ &\qquad\qquad &$T^{(1)}\,$:&$\beta^{(1)}_i$& $1$\\
$U^{(2)}\,$:&$\alpha^{(2)}_i$& $1$ & &$T^{(2)}\,$:&$\beta^{(2)}_i$& $0$\\
$U^{(3)}\,$:&$\alpha^{(2)}_i$& $1$ & &$T^{(3)}\,$:&$\beta^{(3)}_i$& $0$\\
\end{tabular}
\label{ScalarMasses}
\end{equation}
Note the close correlation between the scalar masses and the corresponding
modular weights of the matter fields given in Eq.~(\ref{ModularWeights})
(where the same choice of moduli fields was made).
The scalar mass spectrum in Eq.~(\ref{ScalarMasses}) is non-universal.
This situation is likely to be an important model-building constraint, given
what we know about needed near-degeneracies in certain low-energy squark and
slepton masses. For example, data on $K^0-\bar K^0$ mixing and leptonic
flavor-changing decays like $\mu\to e\gamma$ strongly constrain the mass
differences for squarks and sleptons of the first two generations with the same
electric charge but of different flavor \cite{EN}. The scenario which appears
to emerge in string no-scale supergravity seems to explain this
phenomenological requirement naturally: since all light chiral matter fields
usually arise from the twisted sector, one would assign the first two
generations to the second and third sets (with vanishing scalar masses); the
third generation could be assigned to any of the sets.

\subsection{Fermion masses}
Supersymmetry breaking can also induce masses for (non-chiral) fermions in real
representations of the gauge group. These (unnormalized) masses are given by
the following expression~\cite{FKZ}
\begin{equation}
\left(M_f\right)_{IJ}=m_{3/2}\left(G_{IJ}-G_{IJ\bar K}G^{\bar
K}+\coeff{1}{3}G_IG_J\right)\ .
\label{FermionMasses}
\end{equation}
As above, we focus on the model with K\"ahler potential given
in Eq.~(\ref{completeK}). For fermions in the second and third untwisted sets
[$\alpha^{(2,3)}_i$] and in all of the twisted sets [$\beta^{(1,2,3)}_i$]
one has $G_{IJ}\equiv0$ and $\VEV{G_I}=0$, and thus
\begin{equation}
m_{\widehat\alpha^{(2)}_i}=m_{\widehat\alpha^{(3)}_i}=0\ ,\qquad
m_{\widehat\beta^{(1)}_i}=m_{\widehat\beta^{(2)}_i}=
m_{\widehat\beta^{(3)}_i}=0\ .
\label{mfermions0}
\end{equation}
The remaining fields are $S,\tau,\psi_i$. If we make the simplifying assumption
$\VEV{\psi_i}=0$ (\ie, $\VEV{G_{\psi_i}}=0$) one can show that the normalized
fermion mass matrix reduces to
\begin{equation}
\left(M_f\right)_{IJ}=m_{3/2}\ \bordermatrix{
&\widehat S&\widehat\tau&\widehat\psi_j\cr
\widehat S&2/3&-\sqrt{2}/3&0\cr
\widehat\tau&-\sqrt{2}/3&1/3&0\cr
\widehat\psi_i&0&0&\delta_{ij}\cr}\ .
\label{FermionMassMatrix}
\end{equation}
To obtain this result we have made use of the various normalization factors
given in Eqs.~(\ref{redefinitions1},\ref{redefinitions2}). This matrix has
zero determinant, indicating the presence of a massless eigenstate, namely
the goldstino ($\widetilde\eta$). Indeed, from Eq.~(\ref{FermionMassMatrix})
this eigenstate is $\widetilde\eta=(\widehat S+\sqrt{2}\widehat\tau)/\sqrt{3}$,
in agreement with our previous result in Eq.~(\ref{goldstino}) (for
$\VEV{\psi_i}=0$). From Eq.~(\ref{FermionMassMatrix}) it also follows that the
orthogonal linear combination $\widetilde\eta_\perp=(\sqrt{2}\widehat
S-\widehat\tau)/\sqrt{3}$, and all of the $\widehat\psi_i$ get masses equal to
the gravitino mass, \ie,
\begin{equation}
m_{\widetilde\eta_\perp}=m_{3/2}\ ,\qquad m_{\widehat\psi_i}=m_{3/2}\ .
\label{mfermions1}
\end{equation}

\subsection{A consistency check}
In the previous three subsections we have computed all of the supersymmetry
breaking masses, in particular for the model with K\"ahler potential given in
Eq.~(\ref{completeK}). One can then perform a consistency check of result for
$Q$ given in Eq.~(\ref{Qformula}), since we can calculate directly ${\rm
Str}\,{\cal M}^2=\sum_j(-1)^{2j}(2j+1){\cal M}^2_j=
2Qm^2_{3/2}$. The masses of the complex scalars ($j=0$) are given in
Eq.~(\ref{ScalarMasses}) and contribute to the supertrace (in units of
$m^2_{3/2}$) in the amount of $2(n_{U2}+n_{U3}+n_{T1})$. The masses of the
Majorana fermions ($j=1/2$) are given in
Eqs.~(\ref{mfermions0},\ref{mfermions1}) and contribute $-2(1+n_\psi)$, whereas
the Majorana gaugino masses (given in Eq.~(\ref{m1/2})) contribute $-2d_f$.
Finally the gravitino contributes $-4$. Putting it all together gives
$Q=n_{U2}+n_{U3}+n_{T1}-n_\psi-d_f-3$, which is the result found in
Eq.~(\ref{Qformula}) by less direct means.

\subsection{$A$ terms}
The supersymmetry-breaking cubic scalar couplings (or $A$ terms) are contained
in the term $e^G\,K^I\partial_I\ln W$ (plus hermitian conjugate) of the scalar
potential in Eq.~(\ref{GIGI}). The main input required to evaluate these
couplings is the value of $K^I$ for each of the types of untwisted and twisted
states. For the untwisted states these inputs are given in
Eqs.~(\ref{K^Itau},\ref{K^Ialpha}), \ie,
\begin{equation}
K^{\psi_i}=-\psi_i\ ,\quad
K^{\alpha^{(2)}_i}\approx\alpha^{(2)}_i\ ,\quad
K^{\alpha^{(3)}_i}\approx\alpha^{(3)}_i\ .
\label{K^Iu}
\end{equation}
For the twisted states in the model with K\"ahler potential given in
Eq.~(\ref{completeK}), the first twisted set fields [$\beta^{(1)}_i$] have the
same functional dependence as the second and third untwisted set fields
[$\alpha^{(2,3)}_i$], and therefore the result is as above:
$K^{\beta^{(1)}_i}\approx\beta^{(1)}_i$. For the second and third twisted set
fields, an intermediate step in the calculation that yields the result in
Eq.~(\ref{KIKIapprox}) gives
\begin{equation}
K^{\beta^{(2,3)}_i}\approx
{K_{0\tau\bar\tau}K_2-K_{0\bar\tau}K_{2\tau}\over K_{0\tau\bar\tau}K_2}
\,\beta^{(2,3)}_i\ .
\label{K^Ibeta}
\end{equation}
Inserting the values for $K_0,K_2$ (Eq.~(\ref{K0K1K2})) one finds a zero
result to first order, \ie,
\begin{equation}
K^{\beta^{(1)}_i}\approx\beta^{(1)}_i\ ,\quad
K^{\beta^{(2)}_i}\approx0\ ,\quad
K^{\beta^{(3)}_i}\approx0\ .
\label{K^It}
\end{equation}

With the above results one can proceed to compute the $A$ terms for all the
types of cubic couplings given in Eq.~(\ref{couplings}), with the field
identifications given in Eq.~(\ref{ScalarMasses}). The expression to manipulate
is $e^GK^I\partial_I\ln W=e^{G/2}e^{K/2}K^I\partial_I
W=m_{3/2}e^{K/2}K^I\partial_I W$. Since the cubic superpotential does not
depend on $S$ or $\tau$, all we need to do is take derivatives with respect to
the untwisted and twisted matter fields. Each time one such field is removed
from a cubic coupling by the $\partial_I W$ operation, the corresponding $K^I$
factor puts it back in restoring the original coupling, although a coefficient
($0,1,-1$) is picked up in this process. After summing over all fields in a
given cubic coupling, and over all cubic couplings one ends up with
$m_{3/2}\,c\,e^{K/2} W=m_{3/2}\,c\,\widehat W$ where $\widehat
W(\widehat\phi)=e^{K/2}\,W(\phi)$ is the superpotential written in terms of the
properly normalized fields, as discussed in Sec.~\ref{Field}. The constant $c$
is common to all of the types of cubic couplings in Eq.~(\ref{couplings}), and
in fact $c=1$, \ie,
\begin{equation}
\begin{tabular}{rr}
$\psi\alpha^{(2)}\alpha^{(3)}$ : &$-1+1+1=1$\\
$\psi\beta^{(1)}\beta^{(1)}$ : &$-1+1+1=1$\\
$\alpha^{(2)}\beta^{(2)}\beta^{(2)}$ : &$1+0+0=1$\\
$\alpha^{(3)}\beta^{(3)}\beta^{(3)}$ : &$1+0+0=1$\\
$\beta^{(1)}\beta^{(2)}\beta^{(3)}$ : &$1+0+0=1$\\
\end{tabular}
\label{Aterms}
\end{equation}
Thus we conclude that for all cubic couplings
\begin{equation}
A=m_{3/2}\ .
\label{A}
\end{equation}
(In passing we note that in the old no-scale models, $K^I=\{-[T+\bar T-C_i\bar
C_i],\vec0\}$, and therefore $A\equiv0$.)

\subsection{$\mu$ and $B$}
The possible origin of the low-energy Higgs mixing parameter $\mu$ (and its
associated supersymmetry-breaking bilinear coupling $B$) has been discussed
in the literature for some time. It is well-known that this term ($\mu h_1h_2$)
must be present in the superpotential, and have a magnitude comparable to
all other dimensional parameters of the low-energy theory. In the framework of
string theory, where explicit mass parameters are not present in the
superpotential, the nature of the $\mu$ term is particularly intriguing. Three
scenarios have been put forward:
\begin{itemize}
\item The low-energy theory possesses an additional singlet field ($N$) which
couples to the two Higgs doublets ($\lambda Nh_1h_2$) and gets a vacuum
expectation value which effectively produces $\mu=\lambda\VEV{N}$
\cite{singlet}. Even though such couplings proliferate in fermionic string
models, in all known instances the singlet fields are heavy and decouple from
the low-energy spectrum.
\item The quadratic $\mu$ term arises as an effective non-renormalizable
fourth- (or higher) order term in the superpotential, \ie, ${1\over
M}\lambda_4H\bar Hh_1h_2$ where $M\sim10^{18}\GeV$ is the string scale
\cite{nonren}. In this case $\mu={1\over M}\lambda_4\VEV{H\bar H}$; for
$\mu\sim1\TeV$, one requires $\VEV{H\bar H}^{1/2}\sim10^{11}\GeV$ which is
typical of hidden sector matter condensates in string models.
\item The quadratic $\mu$ term is built into the theory through the K\"ahler
potential, and becomes non-zero and of ${\cal O}(m_{3/2})$ upon supersymmetry
breaking \cite{oldUT,kl,FKZ}.
\end{itemize}

Let us first address the third scenario. From the calculation of the fermion
masses in Eq.~(\ref{mfermions1}) ($m_{\widehat\psi_i}=m_{3/2}$) one could think
that these may come from a superpotential $\mu$ term, \eg,
$\mu\widehat\psi\widehat\psi$ with $\mu={1\over2}m_{3/2}$. However,
Eq.~(\ref{mphipsi}) shows that the corresponding scalar masses vanish
($\widetilde m_{\widehat\psi_i}=0$), a result apparently inconsistent with the
possible presence of a superpotential $\mu$ term. It turns out that things are
more intricate and the interpretation of a $\mu$ term is not inconsistent. What
happens is that one has to split the K\"ahler function into two pieces, one
which is absorbed into the superpotential to provide the $\mu$ term as
suggested in Refs.~\cite{oldUT,kl,FKZ}, and another one which remains as part
of the K\"ahler function. The resolution to the puzzle comes from realizing
that these two pieces give equal and opposite contributions to the squared
scalar masses (but not to the fermion masses). Thus, such a K\"ahler-induced
$\mu$-term does break supersymmetry, even though it can be incorporated into
the superpotential.

Let us illuminate this result by studying a simple example in detail.
Consider the K\"ahler potential $K=-\ln[(\tau+\bar\tau)^2-(\psi+\bar\psi)^2]$,
which gives $\widetilde m_{\widehat\psi}=0$ and $m_{\widehat\psi}=m_{3/2}$. Let
us expand the K\"ahler function to first order in the $\psi$ field
\begin{equation}
K\approx-\ln(\tau+\bar\tau)^2+{2\over(\tau+\bar\tau)^2}\,\psi\bar\psi
+{1\over(\tau+\bar\tau)^2}\,(\psi\psi+\bar\psi\bar\psi)\ .
\label{Example}
\end{equation}
Ignoring the last term in this expression (to be absorbed into $W$) the scalar
mass can be obtained from Eq.~(\ref{betamass}) with $K_0=-\ln(\tau+\bar\tau)^2$
and $K_2=2/(\tau+\bar\tau)^2$. The result is
$\widetilde m^2_{\widehat\psi}=-m^2_{3/2}$. The last term in
Eq.~(\ref{Example}) can be lumped with the superpotential $W\to
We^{K_1\psi\psi}\approx W+WK_1\psi\psi$ with $K_1=1/(\tau+\bar\tau)^2$. Proper
normalization entails multiplying W times
$e^{K/2}$, thus giving the new superpotential term
$e^{K/2}WK_1\psi\psi={1\over2}m_{3/2}\widehat\psi\widehat\psi$, where we have
also properly normalized the $\psi$ field.\footnote{This procedure can be
easily generalized to what would be the case of interest with
$\psi\psi\to\psi_1\psi_2$, as discussed in Ref.~\cite{BIM}. In $SU(5)\times
U(1)$ free-fermionic models one finds $\psi_1={1\over\sqrt{2}}(h_1+h_2)$,
$\psi_2={-i\over\sqrt{2}}(h_1-h_2)$ with $h_1\,(h_2)$ a \r{5}\,(\rb{5}) of
$SU(5)$ \cite{LNY94}. It follows that
$(\psi_1+\bar\psi_1)^2+(\psi_2+\bar\psi_2)^2=2h_1h_2+2(h_1h_2)^*+2h_1h^*_1
+2h_2h^*_2$ and $K_1=K_2=2/(\tau+\bar\tau)^2$. The new superpotential term
is $\mu\hat h_1\hat h_2$ with $\mu=m_{3/2}$, which gives $\widetilde
m_{\widehat h_1}=\widetilde m_{\widehat h_2}=0$ and $m_{\widehat h_1\widehat
h_2}=m_{3/2}$.} We therefore get $\mu={1\over2}m_{3/2}$, which leads to a
superpotential fermion mass $m_{\widehat\psi}=2\mu=m_{3/2}$. This also entails
a superpotential scalar mass-squared $\widetilde
m^2_{\widehat\psi}=4\mu^2=m^2_{3/2}$, which when added to the K\"ahler
potential mass-squared found above gives the expected vanishing
result.\footnote{Note that in principle the coefficients $K_1,K_2$ could be
related in a different manner, with even the $K_2$ piece leading to a vanishing
scalar mass (if $K_2\propto1/(\tau+\bar\tau)$) and to the so-called
``super-soft" supersymmetry breaking terms \cite{Ellwanger} which do not break
supersymmetry.}

If this mechanism for generation of the $\mu$ term is present in realistic
string models, one has to be careful with its embedding into the traditional
supergravity-induced soft-supersymmetry-breaking parameters. For the squared
scalar masses of the Higgs doublets one normally writes $(m_1^2+\mu^2)|H_1|^2
+(m^2_2+\mu^2)|H_2|^2$. The K\"ahler-induced $\mu$ term has the property that
$m^2_1=m^2_2=-\mu^2$, which is what would need to be used as initial conditions
in the corresponding renormalization group equations. (In practice we do not
find this kind of $\mu$ term present in realistic string models, at least
involving the two Higgs doublets which are to remain in the light spectrum.)

Now let us consider the second mechanism for generating a $\mu$ term, namely
via a non-renormalizable coupling in the superpotential. Given such coupling
we would like to know what is the associated supersymmetry breaking $B$ term.
Essentially this term arises in a very similar manner as the $A$ terms
discussed above, that is, from the $e^GK^I\partial_I\ln W$ term in the scalar
potential. There is one crucial difference: the non-renormalizable couplings
may need to be multiplied by powers of the Dedekind eta function to restore
modular invariance of the superpotential, as discussed after
Eq.~(\ref{ModularWeights}). This possible new dependence on the moduli must be
taken into account when taken the derivatives of $W$. Let us first list the
types of quartic and quintic superpotential couplings in free-fermionic models.
In the same spirit as the cubic couplings given in Eq.~(\ref{couplings}), the
types of higher-order couplings can be deduced from Ref.~\cite{KLN},
\begin{equation}
\begin{tabular}{cccc}
Quartic couplings&\qquad\qquad&\multicolumn{2}{c}{Quintic couplings}\\
$ \eta^0\,[T^{(1)}]^2\,[T^{(2)}]^2$&
&$\eta^0\,[T^{(1)}]^2\,[T^{(2)}]^2\,U^{(3)}$
&$\eta^0\,[T^{(1)}]^3\,T^{(2)}\,T^{(3)}$\\
$\eta^0\,[T^{(1)}]^2\,[T^{(3)}]^2$&
&$\eta^0\,[T^{(1)}]^2\,[T^{(3)}]^2\,U^{(2)}$&
$\eta^2\,[T^{(2)}]^3\,T^{(1)}\,T^{(3)}$\\
$\eta^2\,[T^{(2)}]^2\,[T^{(3)}]^2$&
&$\eta^4\,[T^{(2)}]^2\,[T^{(3)}]^2\,U^{(1)}$
&$\eta^2\,[T^{(3)}]^3\,T^{(1)}\,T^{(2)}$\\
\end{tabular}
\label{NRTs}
\end{equation}
Next we need to determine if there is a modular weight imbalance, but this
can only be done in a specific class of models, such as the ones with K\"ahler
potential given in Eq.~(\ref{completeK}). The modular weights are then given
in Eq.~(\ref{ModularWeights}). For every unit of modular weight imbalance
we multiply the non-renormalizable coupling by $\eta^2(\tau)$, as indicated
in Eq.~(\ref{NRTs}). Note that in some instances there is no modular weight
imbalance. The value of the corresponding $B$ parameters can then be determined
in analogy with the procedure followed for the $A$ terms, except when a power
of $\eta^p$ is present. In this case one adds to the result the quantity
\begin{equation}
-(\tau+\bar\tau)(\partial_\tau\eta^p)/\eta^p
=p/2+(\tau+\bar\tau)(p/4\pi)\widehat G_2(\tau)\ .
\label{AddOn}
\end{equation}
With the use of Eqs.~(\ref{K^Iu},\ref{K^It}) one can determine the $B$
parameters for each of the types of quartic or quintic couplings, as follows
(in units of $m_{3/2}$)
\begin{equation}
\begin{tabular}{cc}
Quartic couplings&$B$\\
$[\beta^{(1)}]^2\,[\beta^{(2)}]^2$
&$2$\\
$[\beta^{(1)}]^2\,[\beta^{(3)}]^2$
&$2$\\
$[\beta^{(2)}]^2\,[\beta^{(3)}]^2$
&$1+{\tau+\bar\tau\over2\pi}\,\widehat G_2(\tau)$\\
\end{tabular}
\label{QuarticBs}
\end{equation}
\begin{equation}
\begin{tabular}{lccc}
Quintic couplings&$B$&Quintic couplings&$B$\\
$[\beta^{(1)}]^2\,[\beta^{(2)}]^2\,\alpha^{(3)}$
&$3$
&$[\beta^{(1)}]^3\,\beta^{(2)}\,\beta^{(3)}$
&$3$\\
$[\beta^{(1)}]^2\,[\beta^{(3)}]^2\,\alpha^{(2)}$
&$3$
&$[\beta^{(2)}]^3\,\beta^{(1)}\,\beta^{(3)}$
&$2+{\tau+\bar\tau\over2\pi}\,\widehat G_2(\tau)$\\
$[\beta^{(2)}]^2\,[\beta^{(3)}]^2\,\psi$
&$1+{\tau+\bar\tau\over\pi}\,\widehat G_2(\tau)$
&$[\beta^{(3)}]^3\,\beta^{(1)}\,\beta^{(2)}$
&$2+{\tau+\bar\tau\over2\pi}\,\widehat G_2(\tau)$\\
\end{tabular}
\label{QuinticBs}
\end{equation}
Concerning the calculated values of the $B$ terms, we should note that
$\widehat
G_2(\tau)\approx{\pi^2\over3}(1-24e^{-2\pi\tau})-2\pi/(\tau+\bar\tau)$
with $\widehat G_2(\tau=1)=0$ \cite{Cvetic}. Also, for values of $\tau$ in the
fundamental domain (\ie, $\tau\ge1$ if $\tau$ is real) $\widehat
G_2(\tau)\ge0$. Therefore, the integer values of $B$ shown in
Eqs.~(\ref{QuarticBs},\ref{QuinticBs}) are actually the minimum possible
values.

\section{Field normalizations and Yukawa couplings}
\label{Field}
The superpotential couplings in free-fermionic models are easily calculable
in the string basis. Moreover, in specific models the fermion Yukawa couplings
have led to structures which bear close resemblance to the observed
hierarchical fermion mass spectrum. On the other hand, results about the Yukawa
couplings in the string model are not necessarily directly related to those
which would be observed at low energies. The possible snag lies in the
normalization of the fields in the supergravity lagrangian. For the scalar
fields the relevant term is
\begin{equation}
K_{ij}\partial_\mu\phi_i\partial^\mu\bar\phi_j\ .
\label{kinetic-terms}
\end{equation}
If the K\"ahler potential ($K$) is non-trivial, then the scalar fields would
need to be normalized appropriately. If this effect propagates to the Yukawa
couplings, the physical ones may differ from those naively expected.

We seek a matrix $A$ such that
\begin{equation}
\phi=A\widehat\phi\ ,\qquad \bar\phi=\bar A\widehat{\bar\phi}\ ,
\label{Adef}
\end{equation}
where $\widehat\phi,\widehat{\bar\phi}$ are the properly normalized fields.
{}From the condition $\partial_\mu\phi^T
K\partial^\mu\bar\phi=\partial_\mu\widehat\phi^TA^TK\bar
A\partial^\mu\widehat{\bar\phi}=\partial_\mu\widehat\phi^T
\partial^\mu\widehat{\bar\phi}$ we obtain
\begin{equation}
A=(K^{-1/2})^T\ ,\qquad \bar A=K^{-1/2}\ .
\label{AA}
\end{equation}
Here $K^{-1/2}$ is the matrix obtained by taking the square root of
$K^{M\bar N}$, where $K^{M\bar N}$ has been used above in calculating,
\eg, $K^I=K^{I\bar J}K_{\bar J}$. If $K^{-1/2}$ is obtained, one can determine
the physical Yukawa couplings (which couple properly normalized fields)
from the expression \cite{BIM,FKZ}
\begin{equation}
\widehat\lambda_{ijk}=e^{K/2}\,(K^{-1/2})_{ii'}\,(K^{-1/2})_{jj'}\,
(K^{-1/2})_{kk'}\,\lambda_{i'j'k'}\ ,
\label{yuks}
\end{equation}
where $\lambda_{ijk}$ are the couplings which appear in the superpotential of
the string model. That is, $\widehat W(\widehat\phi)=e^{K/2}\,W(\phi)$.

Calculating the square root of a matrix can be a complicated task. Before we
attempt this, it is quite illuminating to consider the limit of small matter
field vevs. To leading order we only keep the diagonal contributions to
$K_{M\bar N}$, and obtain for the untwisted matter fields
\begin{eqnarray}
&\tau_1,\tau_2\ {\rm set:}\qquad  &[K^{-1/2}]^{(\tau_1,\tau_2)}
\approx{\rm diag}\,[(\tau_1+\bar\tau_1),(\tau_2+\bar\tau_2),\nonumber\\
&&\qquad\qquad\qquad\qquad\qquad\coeff{1}{\sqrt{2}}(\tau_1+\bar\tau_1)^{1/2}(\tau_2+\bar
\tau_2)^{1/2}{\bf1}_{n_\phi}]\ ,\label{K-1/2TU}\\
&\tau\ {\rm set:}\qquad  &[K^{-1/2}]^{(\tau)}
\approx{\rm diag}\,[\coeff{1}{\sqrt{2}}(\tau+\bar\tau),
\coeff{1}{\sqrt{2}}(\tau+\bar\tau){\bf1}_{n_\psi}]
\ ,\label{K-1/2tau}\\
&{\rm \alpha\ set:}\qquad  &[K^{-1/2}]^{(\alpha)}\approx{\rm
diag}\,[{\bf1}_{n_I}]\ ,\label{K-1/2alpha}
\end{eqnarray}
where ${\bf1}_n$ is a vector with $n$ unit entries. In the case of the model
with K\"ahler potential given in Eq.~(\ref{completeK}), and in this same
approximation, we obtain the explicit normalized fields as follows
\begin{eqnarray}
\widehat S&=&{S\over\VEV{S+\bar S}}\ ,\quad
\widehat\tau={\sqrt{2}\,\tau\over\VEV{\tau+\bar\tau}}\ ,\quad
\widehat\psi_i={\sqrt{2}\,\psi_i\over\VEV{\tau+\bar\tau}}\ ;
\label{normalizations1}
\\
\widehat\alpha^{(2)}_i&=&\alpha^{(2)}_i\ ,\quad
\widehat\alpha^{(3)}_i=\alpha^{(3)}_i\ ,\quad
\widehat\beta^{(1)}_i=\beta^{(1)}_i\ ;
\label{normalizations2}
\\
\widehat\beta^{(2)}_i&=&{\beta^{(2)}_i\over\VEV{\tau+\bar\tau}^{1/2}}\ ,
\qquad
\widehat\beta^{(3)}_i={\beta^{(3)}_i\over\VEV{\tau+\bar\tau}^{1/2}}\ .
\label{normalizations3}
\end{eqnarray}

Now let us determine the properly normalized cubic Yukawa couplings. From
Eq.~(\ref{completeK}) we have $\VEV{e^{K/2}}=1/[(S+\bar
S)^{1/2}(\tau+\bar\tau)]$. For the various types of cubic couplings
in Eq.~(\ref{couplings}) we write generically $\widehat\lambda=f\lambda$,
with the normalization factor $f$ given by
\begin{equation}
\begin{tabular}{rr}
$\psi\alpha^{(2)}\alpha^{(3)}$ :
&$e^{K/2}\,\coeff{1}{\sqrt{2}}(\tau+\bar\tau)\,(1)\,(1)=\coeff{1}{2}g$\\
$\psi\beta^{(1)}\beta^{(1)}$ :
&$e^{K/2}\,\coeff{1}{\sqrt{2}}(\tau+\bar\tau)\,(1)\,(1)=\coeff{1}{2}g$\\
$\alpha^{(2)}\beta^{(2)}\beta^{(2)}$ :
&$e^{K/2}\,(1)\,(\tau+\bar\tau)^{1/2}\,(\tau+\bar\tau)^{1/2}=\coeff{1}{\sqrt{2}}g$\\
$\alpha^{(3)}\beta^{(3)}\beta^{(3)}$ :
&$e^{K/2}\,(1)\,(\tau+\bar\tau)^{1/2}\,(\tau+\bar\tau)^{1/2}=\coeff{1}{\sqrt{2}}g$\\
$\beta^{(1)}\beta^{(2)}\beta^{(3)}$ :
&$e^{K/2}\,(1)\,(\tau+\bar\tau)^{1/2}\,(\tau+\bar\tau)^{1/2}=\coeff{1}{\sqrt{2}}g$
\end{tabular}
\label{NormYuks}
\end{equation}
where we have used the result $g^2=1/{\rm Re}\, S$.  We conclude that all the
normalized cubic Yukawa couplings are independent of the moduli,\footnote{This
property does not appear to allow the dynamical determination of Yukawa
couplings via the no-scale mechanism, as recently advocated \cite{KPZ,Dudas}.}
but depend on the dilaton (or the gauge coupling).

Our assumption that the $K^{-1/2}$ matrices are nearly diagonal can be
justified by studying some simple cases where the calculation can be done
exactly. For instance, let us take a $\tau$ set with $n_\psi=1$, which
amounts to a $2\times2$ matrix. The two matrices of relevance are
\begin{equation}
K_{M\bar N}=\coeff{2}{X^2}\left(
\begin{array}{cc}a^2+b^2&-4ab\\-4ab&a^2+b^2\end{array}\right)\ ,
\qquad
K^{-1}=\coeff{1}{2}\left(
\begin{array}{cc}a^2+b^2&2ab\\2ab&a^2+b^2\end{array}\right)\ ,
\label{2x2}
\end{equation}
where $a=\tau+\bar\tau$, $b=\psi+\bar\psi$, and $X=a^2-b^2$. From $K^{-1}$
we can compute the square root,
\begin{equation}
K^{-1/2}=\coeff{1}{\sqrt{2}}\left(
\begin{array}{cc}a&b\\b&a\end{array}\right)\ ,
\label{K-1/2}
\end{equation}
and therefore
\begin{equation}
\left(\begin{array}{c}\tau\\ \psi\end{array}\right)=
\coeff{1}{\sqrt{2}}\left(
\begin{array}{cc}\tau+\bar\tau&\psi+\bar\psi\\ \psi+\bar\psi&\tau+\bar\tau
\end{array}\right)\left(\begin{array}{c}\widehat\tau\\
\widehat\psi\end{array}\right)\ ,
\label{normalized}
\end{equation}
which agrees with Eq.~(\ref{K-1/2tau}) in the limit
$\VEV{\psi+\bar\psi}/\VEV{\tau+\bar\tau}\to0$. However, the exact expression
allows us to study the possibility of {\em moduli-matter} mixing. For the
all-untwisted superpotential coupling
$\lambda_{\psi\alpha\alpha}\psi\alpha\alpha$, we
find in an obvious notation
\begin{eqnarray}
\widehat\lambda_{\widehat\psi\widehat\alpha\widehat\alpha}&=&
{\coeff{1}{\sqrt{2}}(\tau+\bar\tau)\,\lambda_{\psi\alpha\alpha}
\over (S+\bar S)^{1/2}[(\tau+\bar\tau)^2-(\psi+\bar\psi)^2]^{1/2}}\ ,\\
\label{paa}
\widehat\lambda_{\widehat\tau\widehat\alpha\widehat\alpha}&=&
{\coeff{1}{\sqrt{2}}(\psi+\bar\psi)\,\lambda_{\psi\alpha\alpha}
\over (S+\bar S)^{1/2}[(\tau+\bar\tau)^2-(\psi+\bar\psi)^2]^{1/2}}\ .
\label{taa}
\end{eqnarray}
The novelty here is a new (although small) Yukawa coupling between matter
($\widehat\alpha$) and moduli ($\widehat\tau$) fields, of order
$\widehat\lambda_{\widehat\tau\widehat\alpha\widehat\alpha}
={\cal O}[(\psi+\bar\psi)/(\tau+\bar\tau)]$. Otherwise, the results derived
above in the diagonal approximation are quite accurate.

The above exact calculation for $n_\psi=1$ does not allow quantification of
possible {\em matter-matter} mixing through field normalizations. One can
repeat the exercise for $n_\psi=2$ to study the magnitude of such mixings. The
square root of such $3\times3$ matrix can be readily obtained by the use of
Sylvester's formula \cite{Sylvester}
\begin{equation}
P(U)=\sum_{r=1}^n P(\lambda_r)\prod_{j\not=r}{\lambda_j
I-U\over\lambda_j-\lambda_r}\ ,
\label{Sylvester}
\end{equation}
where $U$ is the given matrix, $P$ is the required operation (square root
in our case), and $\lambda_r$ are the eigenvalues of $U$. Mathematica
can be programmed to calculate $K^{-1/2}$ using Sylvester's formula, but the
result is messy. In the limit of interest we find, for example
\begin{equation}
\widehat\lambda_{\widehat\psi_1\widehat\alpha\widehat\alpha}\approx
{\coeff{1}{\sqrt{2}}\lambda_{\psi_1\alpha\alpha}\over(S+\bar S)^{1/2}}
+{\coeff{1}{\sqrt{2}}\lambda_{\psi_2\alpha\alpha}\over(S+\bar S)^{1/2}}
\left[{\coeff{1}{\sqrt{2}}(\psi_1+\bar\psi_1)\coeff{1}{\sqrt{2}}(\psi_2+\bar\psi_2)\over(\tau+\bar\tau)^2}\right]^2\ .
\label{mm}
\end{equation}
The possible matter-matter mixing is therefore very small
${\cal O}[(\psi+\bar\psi)/(\tau+\bar\tau)]^4$. Matter-matter mixing
through the K\"ahler potential (as in the case of the $\mu$ term) or through
the superpotential are therefore the realistic possible sources of mixing.
For $n_\psi=2$ the matter-moduli mixing is also highly suppressed ${\cal
O}[(\psi+\bar\psi)/(\tau+\bar\tau)]^2$.

\section{Possible realistic models}
\label{Possible}
In the previous sections we have explored what consequences would string
no-scale supegravity models have regarding the soft-supersymmetry-breaking
terms and the low-energy Yukawa couplings. However, it remains to be shown
that such models actually exist, \ie, that the first postulate of string
no-scale supergravity is satisfied. Here we describe a search for such models,
and the properties of any that may be found.

\subsection{The search for models}
We have performed a computerized search of free-fermionic string models with
the desired properties. The two cases in Eq.~(\ref{V0=0}) indicate that we
should look for models which {\em effectively} possess one untwisted set with
moduli (a $\tau_1,\tau_2$ set or a $\tau$ set) and the other two untwisted sets
with all moduli projected out. As discussed in Sec.~\ref{Generalities}, this
determination should be done after the cubic superpotential of the model is
calculated, since we would discard ``moduli" which have superpotential
couplings.\footnote{Note that even if a field appears in the superpotential, it
may still be a flat direction if the fields coupled to it conspire in the
appropriate way. In what follows we disregard such exceptional possibilities.}

Let us consider the kind of truncations of the moduli space which could occur
if moduli have superpotential couplings. If we start off with a $\tau$ set,
all that can happen is that the modulus field appears in $W$, and therefore
the set would effectively become an $\alpha$ set. In the case of a
$\tau_1,\tau_2$ set, if both fields appear in $W$ we have an $\alpha$ set,
whereas if only one appears we have a $\tau$ set. In the latter case
($\tau_1,\tau_2\,{\rm set}\to\tau\,{\rm set}$) we would perform the field
redefinition in Eq.~(\ref{taudef}), instead of that in Eq.~(\ref{TUdef}).

With the restriction that we obey Eqs.~(\ref{V0=0}) after moduli present in $W$
are discarded, we have performed a search for free-fermionic models following
the methods described in Ref.~\cite{search}. A free-fermionic model is
specified by a basis of $n$ basis vectors of boundary conditions, plus an
$n\times n$ matrix of GSO projections (the ``$k$-matrix").  Our search is based
on the reasonable assumption that the basis vectors of the fermionic model
contain five ``standard" vectors which have appeared in all known models of
this kind, these are denoted by ${\bf1},S,b_1,b_2,b_3$. Since we are interested
in models with $SU(5)\times U(1)$ observable gauge symmetry, we also assume the
presence of two other vectors ($b_4,b_5$) which have been used in the two
$SU(5)\times U(1)$ string models in the literature \cite{revamped,search}. The
eighth and last vector, called $\alpha$, is decisive. In Ref.~\cite{search}
this vector was allowed to take all possible values consistent with the
free-fermionic model-building rules. In addition, the $k$-matrix was varied
at random. This search was specifically focused on finding models which allow
unification of the low-energy gauge couplings at the string scale. As such,
the model had to contain five \r{10}'s and two \rb{10}'s of $SU(5)$ (a ``5/2
model"), as opposed to the original ``revamped" model which is a ``4/1 model".
In fact, one such 5/2 model was found, which we will call the ``search" model.
The ``search" model is in fact an example of a class of 5/2 models, with
slightly varying properties.

Our purpose here is somewhat different, since the search is more constrained.
We seek either 4/1 or 5/2 models with a single {\em effective} untwisted
$\tau_1,\tau_2$ or $\tau$ set (and two effective untwisted $\alpha$ sets). Our
search procedure consists of picking representative $\alpha$ vectors and
varying the $k$-matrix at random.

\begin{itemize}
\item $\alpha=\alpha_{\rm search}$. In 10,000 $k$-matrices we find $\sim6\%$
$N=1$ supersymmetric models, of which 19 are 5/2 models and none are 4/1
models. All these models possess two $\tau_1,\tau_2$ sets and one $\alpha$ set,
since untwisted (Neveu-Schwarz) states like the moduli depend only on the
choice of basis vectors, and not on the $k$-matrix. Calculation of the cubic
superpotential reveals that the 5/2 models  divide into two ``Yukawa sets":
1/3/2 and 2/3/2,\footnote{An ``$m/n/p$ Yukawa set" includes $m$ potential
up-quark like Yukawa couplings, $n$ potential down-quark like Yukawa couplings,
and $p$ potential charged-lepton Yukawa couplings \cite{search}.} with the
2/3/2 case preferred phenomenologically \cite{search}. Moreover, we discover
that {\em all} models with the 1/3/2 Yukawa set have their $\tau_1,\tau_2$ sets
broken down to $\tau$ sets, whereas {\em all} the models with the preferred
2/3/2 Yukawa set have one $\tau_1,\tau_2$ set broken to an $\alpha$ set and the
other $\tau_1,\tau_2$ set broken to a $\tau$ set. Therefore, the 2/3/2 models
(like the ``search" model) possess precisely the desired moduli content. (The
1/3/2 models give $V_0>0$.) Schematically
\begin{equation}
\begin{array}{c}
{\rm 5/2\ models}\\ \alpha=\alpha_{\rm search}\end{array}
\qquad\left\{
\begin{tabular}{ccc}
Yukawa sets& Moduli&$V_0$\\
1/3/2& $\left\{\begin{array}{l}\tau_1\tau_2\to\tau\\ \tau_1\tau_2\to\tau\\
\alpha\to\alpha\end{array}\right.$&$V_0>0$\\
&&\\
2/3/2& $\left\{\begin{array}{l}\tau_1\tau_2\to\tau\\ \tau_1\tau_2\to\alpha\\
\alpha\to\alpha\end{array}\right.$&$V_0=0$\\
\end{tabular}
\right.
\label{5/2}
\end{equation}
\item $\alpha=\alpha_{\rm revamped}$. In 5,000 $k$-matrices we find 41 4/1
models and no 5/2 models. All these models contain one $\tau_1,\tau_2$ set and
two $\tau$ sets. The 4/1 models come with two possible Yukawa sets (1/3/3 and
2/3/3, the latter is preferred), and in {\em all} instances we find that
the $\tau$ sets are unbroken, whereas the $\tau_1,\tau_2$ set is broken to an
$\alpha$ set, \ie,
\begin{equation}
\begin{array}{c}
{\rm 4/1\ models}\\ \alpha=\alpha_{\rm revamped}\end{array}
\qquad\left\{
\begin{tabular}{ccc}
Yukawa sets& Moduli&$V_0$\\
$\begin{array}{c}1/3/3\\ 2/3/3\end{array}$&
$\left\{\begin{array}{l}\tau_1\tau_2\to\alpha\\ \tau\to\tau\\
\tau\to\tau\end{array}\right.$&$V_0>0$\\
\end{tabular}
\right.
\label{4/1}
\end{equation}
\item $\alpha=\alpha_{\rm 3a}$. We choose two other $\alpha$ vectors which
belong to ``class 3a" in the notation of Ref.~\cite{search}. Models with
$\alpha$ vectors in this class are expected to be 4/1 models ($\alpha_{\rm
revamped}$ belongs to this class). We obtain the same result as in
Eq.~(\ref{4/1}).
\item $\alpha=\alpha_{\rm price}$. Unlike our previous choices for $\alpha$,
$\alpha_{\rm price}$ (introduced in Ref.~\cite{price}) produces both 4/1 and
5/2 models. In this case the three sets are $\tau$ sets and the three moduli
appear in the superpotential, \ie,
\begin{equation}
\begin{array}{c}
{\rm 5/2,\,4/1\ models}\\ \alpha=\alpha_{\rm price}\end{array}
\qquad\left\{
\begin{tabular}{cc}
Moduli&$V_0$\\
$\begin{array}{l}\tau\to\alpha\\\tau\to\alpha\\\tau\to\alpha\end{array}$
&$V_0<0$\\
\end{tabular}
\right.
\label{price}
\end{equation}
These models are not realistic, but we consider them since we want
to establish a connection between the value of $V_0$ and the 4/1 or 5/2
nature of a model.
\item {\em Change $b_4,b_5$}. We finally allow for changes in the core basis,
in addition to varying $\alpha$. In this case 4/1 and 5/2 models are found,
although very unappealing ones (\eg, with no Yukawa couplings!). Nonetheless,
a sample case yields
\begin{equation}
\begin{array}{c}
{\rm 5/2,\,4/1\ models}\\ {\rm change\ b_4,b_5}\end{array}
\qquad\left\{
\begin{tabular}{cc}
Moduli&$V_0$\\
$\begin{array}{l}\tau\to\tau\\\tau\to\tau\\\tau\to\alpha\end{array}$
&$V_0>0$\\
\end{tabular}
\right.
\label{changeb4b5}
\end{equation}
\end{itemize}
\bigskip

\noindent The above search for models, although limited in extent, provides
support for the following
\begin{equation}
{\rm Conjecture:}\qquad {\rm 4/1\ models\ always\ give}\ V_0\not=0\ .
\label{conjecture}
\end{equation}
This would imply that a {\em necessary} condition for $SU(5)\times U(1)$ string
no-scale supergravity models is a 5/2 field content. This condition is
consistent with the string-theory nature of the model which requires
unification at the string scale, which can be accomplished in a 5/2 model.
Moreover, a realistic model which satisfies the postulates of string no-scale
supergravity already exists, namely the ``search" model of Ref.~\cite{search}.

\subsection{A realistic example}
As we just saw, the ``search" model of Ref.~\cite{search} is a good candidate
for a string no-scale supergravity model, with a single effective untwisted
$\tau$ set and a K\"ahler potential of the general form given in
Eq.~(\ref{completeK}). With this information it should be possible to get a
good idea of what the spectrum of sparticles may look like. First of all, we
note that any model of this kind would be a {\em no-parameter model}
\cite{LNZprep}. That is, the whole supersymmetric spectrum would be
unambigously determined. Indeed, with the ability to compute all
soft-supersymmetry-breaking parameters (including $\mu$ and $B$) in terms of
$m_{3/2}$, the high-energy theory would be determined up to the value of
$m_{3/2}$. At low energies one new parameter arises, namely $\tan\beta$, but
one also has two radiative electroweak symmetry breaking conditions which
therefore allow one to determine $m_{3/2}$ and $\tan\beta$. The no-scale
mechanism can then be used to compute the quantity $C$ in Eq.~(\ref{V1def}).
Thus, if the radiative breaking
conditions can be solved, we would have a complete determination of the
sparticle spectrum. In this exercise the top-quark mass is not an independent
parameter: the top-quark Yukawa coupling at the string scale is a hallmark
prediction of string models \cite{t-paper},\footnote{When considering a Yukawa
coupling at the string scale, care must be taken to include any normalization
factors that may arise, as discussed in Sec.~\ref{Field} especially in
Eq.~(\ref{NormYuks}).} and the value of $\tan\beta$ would be self-consistently
determined by the radiative breaking conditions.

Let us first address the question of the value of $Q$ in the ``search" model of
Ref.~\cite{search}. The formula in Eq.~(\ref{Qformula}) is
$Q=n_{U2}+n_{U3}+n_{T1}-n_\psi-d_f-3$. The gauge group is $SU(5)\times
SO(10)\times SU(4)\times U(1)^6$ which gives $d_f=90$. The number of untwisted
and twisted fields is: $n_\psi=13,n_{U2}=14,n_{U3}=16$ and
$n_{T1}=80,n_{T2}=80,n_{T3}=68$, where we count $p$-dimensional representations
of the gauge group as $p$. Putting it all together gives
\begin{equation}
Q=14+16+80-13-90-3=110-106=4\ ,
\label{Q}
\end{equation}
which is remarkably close to the desired zero result ({\em c.f.}, if all terms
were to be added in magnitude, we are off by 2\%).\footnote{In contrast, the
second paper in Ref.~\cite{Coordinate} presents a model with zero vacuum energy
where $Q=-272$. Also, in the ``revamped" model \cite{revamped}, for the value
of $V_0$ closest to zero (it cannot be exactly zero), we get $Q=-83$.}
Pragmatically speaking $Q\not=0$ and a destabilizing one-loop correction to the
scalar potential is expected. However, given our incomplete knowledge of string
dynamics (\eg, the role played by the anomalous $U_A(1)$) and of additional
contributions to $Q$ from massive string states \cite{FKZ} and string loop
corrections, we are not ready to discard this model hastily. For instance,
if the one-loop corrections to the gaugino masses (see Eq.~(\ref{Ma}))
increased them by $4.4\%$, we would obtain $Q=0$. Such small string one-loop
shifts on the scalar and gaugino masses are expected and quantify our statement
that $Q=4$ is a ``small" number. We will therefore proceed exploring the
manifold observable implications of this model, carrying the $Q=4$ result as a
warning flag.

Considering the spectrum of the ``search" model of Ref.~\cite{search}, the
various relevant observable fields (and the sets they belong to) are as follows
\begin{equation}
\begin{tabular}{lccc}
Set&Untwisted fields&Twisted fields\\
First&$\Phi_0,\Phi_1;\,h_1,\bar h_1$&$F_0,F_1,F_4,\bar F_4$\\
Second&$h_2,\bar h_2$&$F_2,\bar f_2,l^c_2;\,\bar F_5,\bar
f_5,l^c_5$\\
Third&$\Phi_3,\Phi_5;\,h_3,\bar h_3$&$F_3,\bar f_3,l^c_3;\, h_{45},\bar h_{45}$
\end{tabular}
\label{set-states}
\end{equation}
The possible moduli are $\Phi_{0,1,3,5}$, of which all but $\Phi_1$ appear in
the cubic superpotential (given in Eq.~(6.3a) in Ref.~\cite{search}).
Therefore, the first set is a $\tau$ set, whereas the other two are $\alpha$
sets. In Ref.~\cite{search} it was argued that $F_4=\{Q_4,d^c_4,\nu^c_4\}$
should contain the third generation squarks, whereas $F_0,F_1,\bar F_4$
contain either Higgs particles or intermediate scale particles, therefore the
first and second generation squarks and sleptons belong to the second and
third twisted sets. Moreover, the light Higgs boson doublets are located inside
$h_1$ and $\bar h_{45}$, which in the usual notation correspond to $H_1,H_2$
respectively. Contrasting Eq.~(\ref{set-states}) with the general result for
the scalar masses in this class of models in Eq.~(\ref{ScalarMasses}) we obtain
the following spectrum of supersymmetry-breaking scalar masses:
\begin{eqnarray}
&{\rm First\ generation:}&m^2_{Q_1,U_1^c,D_1^c,L_1,E_1^c}=0\
,\label{first}\\
&{\rm Second\ generation:}&m^2_{Q_2,U_2^c,D_2^c,L_2,E_2^c}=0\
,\label{second}\\
&{\rm Third\ generation:}&\left\{
\begin{array}{l}
m^2_{Q_3,D^c_3}=m_{3/2}\\
m^2_{U^c_3,L_3,E^c_3}=0
\end{array}
\right.\ ,\label{third}\\
&{\rm Higgs\ masses:}&m^2_{H_1}=0,\quad m^2_{H_2}=0\ . \label{higgses}
\end{eqnarray}
We also know that the gaugino masses are degenerate $m_{1/2}=m_{3/2}$ (see
Eq.~(\ref{m1/2})), and that the $A$ parameter is universal $A=m_{3/2}$ (see
Eq.~(\ref{A})). The $\mu$ parameter is expected to arise at the quintic level
in the superpotential (no suitable terms exist at the quartic level). Since one
of the light Higgs doublets belongs to the first untwisted set ($h_1$) and the
other one to the third twisted set ($\bar h_{45}$), Eq.~(\ref{NRTs}) singles
out only one possible type of quintic term:
$[\beta^{(2)}]^2\,[\beta^{(3)}]^2\,\psi$. Moreover, from Eq.~(\ref{QuinticBs})
we get $B=[1+{\tau+\bar\tau\over\pi}\,\widehat G_2(\tau)]m_{3/2}$,
which has a minimum at $B=m_{3/2}$. In sum, without identifying the specific
quintic term giving rise to $\mu$, we can nonetheless predict the corresponding
$B$ parameter. Therefore, our no-parameter model reduces in practice to a
one-parameter model until we compute such a quintic term.

An important ingredient in the viability of our candidate model is that the
extra (\r{10},\rb{10}) matter representations have suitable masses to allow
gauge coupling unification at the string scale. In fact, the $Q$ and $D^c$
components of these representations should acquire different masses
($\sim10^{12}\GeV$ and $\sim10^6\GeV$ respectively \cite{LNZI}), but this is
allowed since $SU(5)\times U(1)$ is broken at the string scale. Morever, the
$Q$ mass scale is determined by an effective superpotential mass term, whereas
the $D^c$ mass scale is obtained by working out the eigenvalues of the extended
Higgs triplet mass matrix \cite{search}.

To summarize, the above candidate model has several notable properties: (i)
a $Q$ parameter tantalizingly close to zero, (ii) a benign non-universal
spectrum of supersymmetry-breaking scalar masses which bears close resemblance
to the old ``no-scale" result, (iii) a universal trilinear term $A=m_{3/2}$,
(iv) a $\mu$ parameter at the quintic level of superpotential interactions
associated with $B\ge m_{3/2}$, and (v) extra \r{10},\rb{10} representations
dynamically required and likely to possess the desired mass spectrum. The
low-energy consequences of this model will be explored in detail elsewhere
\cite{LNZprep}.

\section{Conclusions}
\label{Conclusions}
In this paper we have explored the postulates of string no-scale supergravity
in the context of free-fermionic string models. These postulates are not
trivially satisfied, and in fact impose important new restrictions on string
model building. In particular the moduli sector should be rather minimal, and
the massless matter spectrum and gauge group should be correlated in a way
such that the parameter $Q$ vanishes.  We have given plausibility arguments
indicating that this condition may be possible to satisfy in specific models,
and in fact presented a model where $Q$ is very close to zero.
For all (untwisted and twisted) matter fields we have computed explicitly the
associated supersymmetry breaking parameters. Models of this kind are in fact
``no-parameter" models, with all needed masses and couplings completely
determined. A search for free-fermionic models which satisfy the minimal
necessary conditions yielded one candidate model with the $SU(5)\times U(1)$
observable gauge group, calculated novel supersymmetry breaking parameter
space, and several desirable properties regarding the stability of the no-scale
mechanism. This search also appears to imply that viable $SU(5)\times U(1)$
models always contain additional matter representations that allow unification
at the string scale.

It is interesting to remark that the models studied in this paper possess a
goldstino composition with significant admixtures of both ``dilaton" and
``moduli". This hybrid scenario borrows desirable features from the two
extremes, where either one or the other dominates the goldstino field. From the
dilaton admixture we get a tree-level contribution to the gaugino mass, and
from the moduli admixture (or more properly old no-scale) we get
universality of scalar masses in the subset of fields where it is desired.

We should also note that the no-scale supergravity realized in free-fermionic
models [$SO(2,n)/SO(2)\times SO(n)$] differs in structure from that in the old
no-scale models [$SU(n,1)/U(1)\times SU(n)$]. The effects of the different
structures is most evident in the computation of the vacuum energy and in the
computation of the supersymmetry breaking parameters.

Throughout our discussion we have said little about the supersymmetry breaking
mechanism which creates $\VEV{W}\not=0$, thus implicitly assuming that its
precise nature would not affect our results. In gaugino condensation models,
the non-perturbative superpotential depends explicity on $S$ and fixes its
value, but this $S$-dependence is likely to also affect the calculation of
$V_0$. Nonetheless, it may be possible to retain all the desirable no-scale
supergravity properties \cite{FKZ}. In coordinate-dependent compactifications
we expect our results to hold since $W$=constant, although here the question
is: what determines $S$? The no-scale mechanism extended to the $S$ field would
appear to answer this question. A third possibility to obtain $\VEV{W}\not=0$
appears possible in models with an anomalous $U_A(1)$. In this case various
singlet fields would acquire vacuum expectations values $\vev{\phi}/M\sim1/10$
and a cubic term in the superpotential would give
$\VEV{W}=(\vev{\phi}/M)^3\sim10^{-3}$, if the flatness conditions can be
simultaneously satisfied. The crucial question in the phenomenologically
viability of these supersymmetry breaking mechanisms is the whether the
calculated value of $m_{3/2}$ is of electroweak scale size, since all
supersymmetry-breaking parameters are proportional to $m_{3/2}$.

\section*{Acknowledgments}
We would like to thank Costas Kounnas for useful and encouraging discussions
and for reading the manuscript. We would also like to thank Kajia Yuan for
useful discussions at the earlier stages of this project, for reading the
manuscript, and for consultations regarding the twisted sector K\"ahler
potential. This work has been supported in part by DOE grant
DE-FG05-91-ER-40633.

\appendix
\section{Field redefinitions}
\label{Redefinitions}
As discussed in Sec.~\ref{Fermionic}, the modular properties of the theory
are evident in the supergravity basis, while the direct string calculations
are in the string basis. Here we discuss how one goes from one basis to the
other, or more specifically, how Eqs.~(\ref{TUdef}) and (\ref{taudef}) are
obtained from Eq.~(\ref{Kdef}). Our purpose is to identify a transformation
which leaves the K\"ahler function ($G$) invariant. In this way all results
which follow from it will not depend on the transformation. Some discussion
of the relevant transformation has been given in Ref.~\cite{FKPZII}.

The main observation is that the two forms of the untwisted K\"ahler potential
are simply two different parametrizations of the metric for the same coset
space: $SO(2,n)/SO(2)\times SO(n)$. Following Gilmore \cite{Gilmore}, we
introduce complex variables $t_j=x_j+iy_j$ with $1\le j\le n+2$, which describe
the coset space of dimension $2n$ (two of the variables are auxiliary). The
first parametrization is in terms of the variables $\alpha_j$
\begin{equation}
\alpha_j={t_j\over t_{n+1}-it_{n+2}}\ ,\qquad 1\le j\le n\ ,
\label{alphas}
\end{equation}
which have the following properties \cite{Gilmore}
\begin{eqnarray}
\left|\sum_{j=1}^n\alpha^2_j\right|&<&1\ , \label{prop1}\\
Y(\alpha)=
1-2\sum_{j=1}^n\alpha_j\bar\alpha_j+\left|\sum_{j=1}^n\alpha^2_j\right|^2&=&
{4\over|t_{n+1}-it_{n+2}|^2}>0\ .\label{prop2}
\end{eqnarray}
This parametrization corresponds to that given in Ref.~\cite{LNY94} and used in
Eq.~(\ref{Kdef}). In fact, $Y(\alpha)$ is the argument of the logarithm in the
K\"ahler potential. The second parametrization is in terms of the variables
$\beta_k$
\begin{equation}
\beta_k={t_k\over t_1-t_{n+2}}\ ,\qquad 2\le k\le n+1\ ,
\label{betas}
\end{equation}
which have the property
\begin{equation}
Y(\beta)=\sum_{k=2}^n(\beta_k-\bar\beta_k)^2-(\beta_{n+1}-\bar\beta_{n+1})^2=
{4\over|t_{n+2}-t_1|^2}>0\ .
\label{prop}
\end{equation}
Performing the phase transformation $\beta_k\to i\beta_k$, we get
\begin{equation}
Y(\beta)\to
Y(\beta)=(\beta_{n+1}+\bar\beta_{n+1})^2-\sum_{k=2}^n(\beta_k+\bar\beta_k)^2\ ,
\label{Ytau}
\end{equation}
which reproduces Eq.~(\ref{taudef}) with the identifications
$\beta_{n+1}\to\tau$, $\beta_k\to\psi_k$. Moreover, we can rewrite
Eq.~(\ref{Ytau}) as follows
\begin{eqnarray}
Y(\beta)&=&(\beta_{n+1}+\bar\beta_{n+1})^2-(\beta_{n}+\bar\beta_{n})^2-\sum_{k=2}^{n-1}(\beta_k+\bar\beta_k)^2\nonumber\\
&=&(\tau_1+\bar\tau_1)(\tau_2+\bar\tau_2)-\sum_{i=1}^{n_\phi}(\phi_i+\bar\phi_i)^2
\label{YTU}
\end{eqnarray}
where the second equality follows from the identifications
$\beta_{n+1}+\beta_n\to \tau_1$ and $\beta_{n+1}-\beta_n\to \tau_2$. This
result reproduces Eq.~(\ref{TUdef}).

We now show that the transformation $\alpha\leftrightarrow
t\leftrightarrow\beta$ leaves $G$ unchanged. Let us write $e^G$ as follows
\begin{equation}
e^G={|W|^2\over Y_0Y_1Y_2Y_3}={\lambda_{ijk}\lambda^*_{i'j'k'}\over Y_0}\,
{\alpha^{(1)}_i\bar\alpha^{(1)}_{i'}\over Y_1(\alpha)}\,
{\alpha^{(2)}_j\bar\alpha^{(2)}_{j'}\over Y_2(\alpha)}\,
{\alpha^{(3)}_k\bar\alpha^{(3)}_{k'}\over Y_3(\alpha)}\ ,
\label{e^G}
\end{equation}
where $Y_0=(S+\bar S)$, $Y_{1,2,3}=e^{-K_{(1,2,3)}}$ with the $K$'s in
Eq.~(\ref{Kdef}), and the superpotential is
$W=\lambda_{ijk}\alpha^{(1)}_i\alpha^{(2)}_j\alpha^{(3)}_k$. Now we note that
from Eqs.~(\ref{betas}),(\ref{prop}) we can write
\begin{equation}
{\beta_i\bar\beta_{i'}\over Y(\beta)}={t_it_{i'}\over|t_1-t_{n+2}|^2}\cdot
{|t_{n+2}-t_1|^2\over4}=\coeff{1}{4}t_i\bar t_{i'}\ ,
\label{bb}
\end{equation}
whereas from Eqs.~(\ref{alphas}),(\ref{prop2}) we can write
\begin{equation}
{\alpha_i\bar\alpha_{i'}\over
Y(\alpha)}={t_it_{i'}\over|t_{n+1}-it_{n+2}|^2}\cdot
{|t_{n+1}-it_{n+2}|^2\over4}=\coeff{1}{4}t_i\bar t_{i'}\ .
\label{aa}
\end{equation}
Therefore we conclude that
$\beta_i\bar\beta_{i'}/Y(\beta)=\alpha_i\bar\alpha_{i'}/Y(\alpha)$, and thus
Eq.~(\ref{e^G}) shows that $e^G$ remains invariant when written in terms
of the $\beta$ variables. Note that the transformation involves the K\"ahler
potential {\em and} the superpotential, and that the superpotential has the
same couplings when written in terms of the $\beta$ variables. It appears
unnecessary to relate the $\alpha$ to the $\beta$ variables directly (\ie,
eliminating $t$), although this has apparently been done in Ref.~\cite{FKPZII}.

\section{Example of twisted sector K\"ahler potential}
\label{AppB}
The twisted sector contribution to the K\"ahler potential in free-fermionic
models was calculated in Ref.~\cite{oldTS} for a simple model with $N=1$
spacetime supersymmetry and fermionic basis ${\cal
B}=\{{\bf1},S,b_1,b_2,b_3\}$. This model has only three (massless) twisted
sectors: $b_1,b_2,b_3$. The result, obtained to lowest order in the twisted
fields, is
\begin{equation}
K_{\rm TS}=
 \sum_i^{n_{T1}} \beta^{(1)}_i\bar\beta^{(1)}_i\ e^{{1\over2}[K_{(2)}+K_{(3)}]}
+\sum_i^{n_{T2}} \beta^{(2)}_i\bar\beta^{(2)}_i\ e^{{1\over2}[K_{(1)}+K_{(3)}]}
+\sum_i^{n_{T3}} \beta^{(3)}_i\bar\beta^{(3)}_i\ e^{{1\over2}[K_{(1)}+K_{(2)}]}
\label{oldKTS}
\end{equation}
where $\beta^{(1,2,3)}$ are the twisted fields (numbering $n_{T1,T2,T3}$) in
the $b_{1,2,3}$ sectors, and $K_{(1,2,3)}$ are the contributions to the
K\"ahler potential from the untwisted fields as given in Eq.~(\ref{Kdef}).
Realistic free-fermionic models contain more basis vectors and a great deal
more massless twisted sectors. For instance, the ``search" model discussed
above has basis ${\cal B}=\{{\bf1},S,b_1,b_2,b_3,b_4,b_5,\alpha\}$ and 22
massless twisted sectors. Despite this enlargement of the model, we conjecture
that the structure of the twisted sector K\"ahler potential remains as simple
as in Eq.~(\ref{oldKTS}) once we generalize the concept of twisted sector to
``twisted set" with the meaning given in Sec.~\ref{Generalities}. In what
follows we prove this conjecture by explicit calculation in the context of the
``search" model.\footnote{For a more general discussion of twisted sector
K\"ahler potentials in fermionic models see Ref.~\cite{LNY95}.}

We start by listing all of the massless states of the model divided into
untwisted and twisted fields and by the set they belong to
\begin{equation}
\begin{tabular}{lcc}
Set&Untwisted fields&Twisted fields\\
&$\Phi_0,\Phi_1$&$F_0,F_1$\quad {\small$[b_1]$}\\
First&$\Phi_{23},\bar\Phi_{23}$&$F_4,\bar F_4$\\
&$h_1,\bar h_1$&$\F_3,\Fb_{1,2,4},D_{1,2,5,6}$\\
&&\\
&$\eta_1,\bar\eta_1$&$F_2,\bar f_2,l^c_2$\quad {\small$[b_2]$}\\
Second&$\Phi_{31},\bar\Phi_{31}$&$\bar F_5,\bar f_5,l^c_5$\\
&$h_2,\bar h_2$&$\F_{1,5,6},\Fb_3,D_{3,7},T_{1,3}$\\
&&\\
&$\Phi_3,\Phi_5$&$F_3,\bar f_3,l^c_3$\quad {\small$[b_3]$}\\
Third&$\eta_2,\bar\eta_2$&$h_{45},\bar h_{45}$\\
&$\Phi_{12},\bar\Phi_{12}$
&$\phi_{45},\bar\phi_{45},\phi^+,\bar\phi^+,\phi^-,\bar\phi^-,\phi_{3,4},\bar\phi_{3,4}$\\
&$h_3,\bar h_3$&$\F_{2,4},\Fb_{5,6},D_4,T_2$
\end{tabular}
\label{AllStates}
\end{equation}
(The $[b_{1,2,3}]$ that appear next to some states indicates that these states
belong to that particular twisted sector.)
With the exception of $\Phi_{0,1,3,5}$, the above fields carry charges
under $SU(5)\times SU(4)\times SO(10)$ and various $U(1)$ symmetries:
$\Phi_{23,31,12},\bar\Phi_{12,31,12},\eta_{1,2},\bar\eta_{1,2}$ and
$\phi_{45},\bar\phi_{45}$, $\phi^+,\bar\phi^+$, $\phi^-,\bar\phi^-$,
$\phi_{3,4},\bar\phi_{3,4}$ are gauge singlets; $F_{0,1,2,5}$ ($\bar F_{4,5}$)
are \r{10}
(\rb{10}) under $SU(5)$, $\bar f_{2,3,5}$ ($l^c_{2,5,3}$) are \rb{5} (\r{1})
under $SU(5)$; $\F_{1,2,3,4,5,6}$ ($\Fb_{1,2,3,4,5,6}$) are \r{4} (\rb{4})
under $SU(4)$, and $D_{1,2,3,4,5,6,7}$ are \r{6} under $SU(4)$; $T_{1,2,3}$ are
\r{10} under $SO(10)$.

To verify our conjecture (at least to lowest order) we work in the string basis
and expand the exponentials in Eq.~(\ref{oldKTS}) to first order in the
untwisted states:
$K_{(I)}\approx\sum_{i}^{n_I}\alpha^{(I)}_i\bar\alpha^{(I)}_i$. We end up
with generic terms of the form
\begin{equation}
\beta^{(I)}_i\bar\beta^{(I)}_i
+\coeff{1}{2}\beta^{(I)}_i\bar\beta^{(I)}_i\sum_j\left[\alpha^{(J)}_j\bar\alpha^{(J)}_j+\alpha^{(K)}_j\bar\alpha^{(K)}_j\right]\ ,
\label{generic}
\end{equation}
where $J,K\not=I$. To verify the presence of the quartic term with the
$1\over2$ coefficient we need to compute string scattering amplitudes of
the type
$\VEV{\beta^{(I)}_i\bar\beta^{(I)}_i\alpha^{(J)}_j\bar\alpha^{(J)}_j}$,
which should exhibit a term proportional to $s$ (in fact $(g^4/4)\,(s/2)$).
This type of calculations have been performed in
detail in Ref.~\cite{LNY94}. Here we give the new results of interest and point
out subtleties that need to be dealt with in the process.

To verify the old result in Ref.~\cite{oldTS}, we need to pick $\beta$'s from
the twisted sectors $b_{1,2,3}$ (see Eq.~(\ref{AllStates})). For instance,
for $\alpha^{(1)}=\Phi_{23}$ and $\beta^{(1)}=F_{0,1}$ we find
\begin{equation}
\VEV{F_{0,1}\Phi_{23}\Phi_{23}^\dagger F_{0,1}^\dagger}={g^2\over4}\,{su\over
t} \ ,
\label{Ex1}
\end{equation}
which exhibits no term $\propto s$ since both $\alpha$ and $\beta$ belong to
the same set. The $su/t$ term is just the expected graviton exchange
contribution. There are no further contributions since $F_{0,1}$ and
$\Phi_{23}$ do not have any $U(1)$ charges in common (\ie, no ``D terms") or
appear in the same superpotential coupling (\ie, no ``F terms").\footnote{A
listing of all $U(1)$ charges associated with the fields in
Eq.~(\ref{AllStates}) is given in Table~4 in Ref.~\cite{search}; the cubic and
quartic superpotential is given in Eq.~(6.3) of this same reference.} Now let
us consider $\alpha^{(2)}=\Phi_{31},\bar\Phi_{31}$ and $\beta^{(1)}=F_{0,1}$,
\begin{eqnarray}
\VEV{F_{0,1}\Phi_{31}\Phi_{31}^\dagger F_{0,1}^\dagger}&=&
{g^2\over4}\left[{s\over2}+{su\over t}+2\ln2 s\right]
-{g^2\over2}\left({s-u\over t}-1\right)\ ,\label{Ex2a}\\
\VEV{F_{0,1}\bar\Phi_{31}\bar\Phi_{31}^\dagger F_{0,1}^\dagger}&=&
{g^2\over4}\left[{s\over2}+{su\over t}-2\ln2 s\right]
+{g^2\over2}\left({s-u\over t}-1\right)\ .
\label{Ex2b}
\end{eqnarray}
In these expressions we see the expected graviton exchange term ($\propto
su/t$) and also the contact term $(g^2/4)(s/2)$, signaling the non-trivial
quartic coupling in the K\"ahler function (see Eq.~(\ref{generic})). The
last term is just a ``D term": under a common $U(1)$ $F_{0,1}$ carry $-1/2$
charge, whereas $\Phi_{31}\,(\bar\Phi_{31})$ carries $+1\,(-1)$ charge, which
give $D=-{1\over2}|F_{0,1}|^2+|\Phi_{31}|^2-|\bar\Phi_{31}|^2$. Therefore, we
expect gauge boson exchange [$\propto(s-u)/t$] and a contact term from
$-{g^2\over2}D^2$, all evidently present in Eqs.~(\ref{Ex2a},\ref{Ex2b}). We do
not expect an ``F term" since there is no superpotential coupling involving
these fields. The last matter concerns the disturbing term $\propto 2\ln2 s$.
In Ref.~\cite{LNY94} it was shown that the proper coordinates in which to write
the untwisted sector K\"ahler function may be linear combinations of the string
coordinates, in this case $\chi_1={1\over\sqrt{2}}(\Phi_{31}+\bar\Phi_{31})$
and  $\chi_2={-i\over\sqrt{2}}(\Phi_{31}-\bar\Phi_{31})$. The amplitudes to
consider are then $\VEV{F_{0,1}\chi_{1,2}\chi^\dagger_{1,2}F^\dagger_{0,1}}=
{1\over2}\VEV{F_{0,1}\Phi_{31}\Phi_{31}^\dagger F_{0,1}^\dagger}
+{1\over2}\VEV{F_{0,1}\bar\Phi_{31}\bar\Phi_{31}^\dagger F_{0,1}^\dagger}
={g^2\over4}[s/2+su/t]$, which is the expected result.

Continuing in this fashion one can calculate all of the terms of the form
(\ref{generic}), where the $\beta$'s come exclusively from the $b_{1,2,3}$
sectors. In each instance one can account for all pieces of the
amplitudes, thus confirming the old result in Ref.~\cite{oldTS}. We have
conjectured that this result can be extended to all states in the spectrum.
This conjecture has been verified explicitly for all twisted sectors in this
model. For instance, consider $\alpha^{(1)}=\Phi_0$ and
$\beta^{(1)}=F_4$,\footnote{This amplitude involves the Ising model correlator
$\VEV{\sigma f f \sigma}={1\over2}|z_\infty|^{-1/4}[4(1-z)^{-2}+z^{-1}]^{1/2}$,
which has been calculated using the methods of Ref.~\cite{KLN}.}
\begin{equation}
\VEV{F_4\Phi_0\Phi_0^\dagger F_4^\dagger}={g^2\over4}\,{su\over t}-{g^2\over2}
\ .
\label{Ex3}
\end{equation}
According to our conjecture, in this case we do expect a term $\propto s$ since
both fields belong to the first set. We also get the usual graviton exchange
term, and a contact term $-g^2/2$. This latter is an ``F term" originating from
the superpotential coupling ${1\over2}g\sqrt{2}F_4\bar F_4\Phi_0$. Another
example would be to take $\alpha^{(2)}=\Phi_{31},\bar\Phi_{31}$ and
$\beta^{(1)}=F_4$. The result in this case is identical to that in
Eqs.~(\ref{Ex2a},\ref{Ex2b}); repeating the subsequent discussion shows the
appearance of the term $\propto s$, as conjectured. A lot more of these brute
force calculations shows that the conjecture holds for all states in the
massless spectrum. That is, in Eq.~(\ref{oldKTS}) one is to intepret
$\beta^{(I)}$ as fields belonging to the $I$-th set as defined in
Eq.~(\ref{charges}).

We should add that we have also verified the results of Ref.~\cite{LNY94} for
the untwisted sector of the ``search" model. The novelty here is that some of
the untwisted singlet fields (some of the presumed moduli) have superpotential
couplings (in fact $\Phi_{0,3,5}$), a feature that does not alter the structure
of the K\"ahler function derived in Ref.~\cite{LNY94}. We have also repeated
the above exercise (and therefore proven our conjecture) for the ``revamped"
model of Ref.~\cite{revamped}. For remarks regarding the generality of these
results see Ref.~\cite{LNY95}.

In the main text we have used the twisted sector contribution to the K\"ahler
potential to write down Eq.~(\ref{completeK}). In this case we have one
untwisted set which is a $\tau$ set, and the other two are $\alpha$ sets. In
Eq.~(\ref{completeK}) the two $\alpha$ sets have been expanded to first order
in $\alpha^{(2,3)}_i$. The same expansion has been carried out in the exponents
in Eq.~(\ref{oldKTS}), reducing it to unity for the first twisted set,
and to the square-root of the argument of the $\ln$ in the first untwisted set
K\"ahler potential. Equation~(\ref{completeK}) then follows immediately. We
should remark that these various approximations in calculating the K\"ahler
function are immaterial as far as the observable quantities which we have
calculated are concerned.

\end{document}